\documentclass[a4paper,12pt]{article}
\usepackage{amsmath,amssymb,amsthm,epic,eepic}
\usepackage[english]{babel}
\title{Classification \\ of integrable equations on quad-graphs. \\
The consistency approach}
\author{V.E.\,Adler
 \thanks{Landau Institute of Theoretical Physics, 12 Institutsky pr., 
 142432 Chernogolovka, Russia. E--Mail: {\tt adler}@{\tt itp.ac.ru}}
\and A.I.\,Bobenko
 \thanks{Institut f\"ur Mathematik, Technische Universit\"at Berlin,
 Str. des 17. Juni 136, 10623 Berlin, Germany. E--Mail: {\tt
 bobenko}@{\tt math.tu-berlin.de}}
\and Yu.B.\,Suris
 \thanks{Institut f\"ur Mathematik, Technische Universit\"at Berlin,
 Str. des 17. Juni 136, 10623 Berlin, Germany. E--Mail: {\tt
 suris}@{\tt sfb288.math.tu-berlin.de}}}
\date{}

\def\a{\alpha}
\def\b{\beta}
\def\g{\gamma}
\def\d{\delta}
\def\eps{\varepsilon}

\def\({\left(}
\def\){\right)}
\def\<{\langle}
\def\>{\rangle}
\def\wx{\widetilde{x}}
\def\wX{\widetilde{X}}
\def\itbf{\itshape\bfseries}

\newtheorem{theorem}{Theorem}
\newtheorem{lemma}[theorem]{Lemma}
\newtheorem{proposition}[theorem]{Proposition}

\newcommand{\cQ}{{\cal Q}}
\newcommand{\cX}{{\cal X}}

\newcount\xx   \newcount\xy    \newcount\XX   \newcount\XY
\newcount\yx   \newcount\yy    \newcount\YX   \newcount\YY
\newcount\zx   \newcount\zy    \newcount\ZX   \newcount\ZY
\newcount\xyx  \newcount\xyy   \newcount\XYX  \newcount\XYY
\newcount\xzx  \newcount\xzy   \newcount\XZX  \newcount\XZY
\newcount\yzx  \newcount\yzy   \newcount\YZX  \newcount\YZY
\newcount\xyzx \newcount\xyzy  \newcount\XYZX \newcount\XYZY
\newcount\MM \newcount\MD
\newcount\ox \newcount\oy
\def\summ(#1,#2,#3){#1=#2 \advance#1 #3}
\def\calcshift(#1,#2)
 {#1=#2 \multiply#1 by \MM \divide#1 by \MD \advance#1 -#2 \divide#1 2}
\def\calc(#1,#2,#3)
 {#1=#2 \multiply#1 by \MM \divide#1 by \MD \advance#1 -#3}
\def\Sum{
 \summ(\xyx,\xx,\yx)   \summ(\xyy,\xy,\yy)
 \summ(\xzx,\xx,\zx)   \summ(\xzy,\xy,\zy)
 \summ(\yzx,\yx,\zx)   \summ(\yzy,\yy,\zy)
 \summ(\xyzx,\xyx,\zx) \summ(\xyzy,\xyy,\zy)}
\begin{document}
\maketitle
\begin{abstract} A classification of discrete integrable systems
on quad--graphs, i.e. on surface cell decompositions with
quadrilateral faces, is given. The notion of integrability laid in
the basis of the classification is the three--dimensional
consistency. This property yields, among other features, the
existence of the discrete zero curvature with a spectral
parameter. For all integrable systems of the obtained
exhaustive list, the so called three--leg forms are found. This
establishes Lagrangian and symplectic structures for these systems, 
and the connection to discrete systems of the Toda type 
on arbitrary graphs. Generalizations of these ideas to
the three--dimensional integrable systems and to the quantum
context are also discussed.
\end{abstract}

\section{Introduction}\label{sect:intro}

The idea of consistency (or compatibility) is in the core of the
integrable systems theory. One is faced with it already at the
very definition of the complete integrability of a Hamiltonian
flow in the Liouville--Arnold sense, which means exactly that the
flow may be included into a complete family of commuting
(compatible) Hamiltonian flows \cite{A}. Similarly, it is a
characteristic feature of soliton (integrable) partial
differential equations that they appear not separately but are
always organized in hierarchies of commuting (compatible) systems.
It is impossible to list all applications or reincarnations of
this idea. We mention only a couple of them relevant for our
present account. A condition of existence of a number of commuting
systems may be taken as the basis of the {\it symmetry approach}
which is used to single out integrable systems in some general
classes and to classify them \cite{MSY}. With the help of the {\it
Miwa transformation} one can encode a hierarchy of integrable
equations, like the KP one, into a single discrete (difference)
equation \cite{M}. Another way of relating continuous and discrete
systems, connected with the idea of compatibility, is based on the
notion of {\it B\"acklund transformations} and the {\it Bianchi
permutability theorem} for them \cite{Bi}. This notion, born in
the classical differential geometry, found its place in the modern
theory of discrete integrable systems \cite{QNCL}. A
sort of the synthesis of the analytic and the geometric approach
was achieved in \cite{BP1} and is being actively developed since
then, see a review in \cite{BP2}. These studies have revealed the
fundamental importance of discrete integrable systems; it is
nowadays a commonly accepted idea that they may be regarded as the
cornerstone of the whole theory of integrable systems. For
instance, one believes that both the differential geometry of
``integrable'' classes of surfaces and their transformation theory
may be systematically derived from the multidimensional lattices
of consistent B\"acklund transformations \cite{BP2}.

So, the consistency of discrete equations steps to the front stage of the
integrability theater. We say that
\begin{quote}
a $d$--dimensional discrete equation possesses the {\itbf consistency}
 property, if it may be imposed in a consistent way on all
$d$--dimensional sublattices of a $(d+1)$--dimensional lattice
\end{quote}
(a more precise definition will be formulated below). As it is
seen from the above remarks, the idea that this notion is closely
related to integrability, is not new. In the case $d=1$ it was
used as a possible definition of integrability of maps in
\cite{V}. A clear formulation in the case $d=2$ was given recently
in \cite{NW}. A decisive step was made in \cite{BS}: it was shown
there that the integrability in the usual sense of the soliton
theory (as existence of the zero curvature representation) {\it
follows} for two--dimensional systems from the three--dimensional
consistency. So, the latter property may be considered as a
definition of integrability, or its criterion which may be checked
in a completely algorithmic manner starting with no more
information than the equation itself. Moreover, in case when this
criterion gives a positive result, it delivers also the transition
matrices participating in the discrete zero curvature
representation. (Independently, this was found in \cite{N}.)

In the present paper we give a further application of the consistency property:
we show that it provides an effective tool for finding and classifying all
integrable systems in certain classes of equations. We study here integrable
one--field equations on {\it quad--graphs}, i.e. on cellular decompositions of
surfaces with all faces (2--cells) being  quadrilateral \cite{BS}. In Sect.
\ref{sect:theorem} we formulate our main result (Theorem 1), consisting of
a complete classification of integrable systems on quad--graphs. Of course,
we provide also a detailed discussion of the assumptions of Theorem 1.
Sect. \ref{sect:solv1}, \ref{sect:solv2} are devoted to the proof
of the main theorem. In Sect. \ref{sect:3legs} we discuss the so called
three--leg forms \cite{BS} of all integrable equations from Theorem 1.
This device, artificial from the first glance, proves to be extremely useful
in several respects. First, it allows to establish a link with the discrete
Toda type equations introduced in full generality in \cite{A2}. Second,
it provides a mean to the most natural derivation of invariant symplectic
structures for evolution problems generated by equations on quad--graphs.
This is discussed in Sect. \ref{sect:Lagrange}. Further,
Sect. \ref{sect:Backlund} contains a brief discussion of the relation of
the equations listed in Theorem 1 with B\"acklund transformations for
integrable partial differential equations. Finally, in Sect.
\ref{sect:conclusions} we discuss some perspectives for the further work,
in particular the consistency approach for three--dimensional systems, and
discrete quantum systems.

\section{Formulation of the problem; results}\label{sect:theorem}
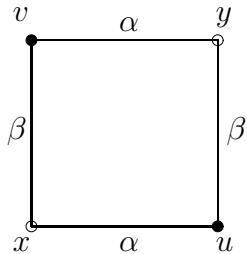
\begin{figure}[htbp]
\begin{center}
\setlength{\unitlength}{0.06em}
\begin{picture}(200,140)(-50,-20)
  \put(100,  0){\circle*{6}} \put(0  ,100){\circle*{6}}
  \put(  0,  0){\circle{6}}  \put(100,100){\circle{6}}
  \put( 0,  0){\line(1,0){100}}
  \put( 0,100){\line(1,0){100}}
  \put(  0, 0){\line(0,1){100}}
  \put(100, 0){\line(0,1){100}}
  \put(-10,-13){$x$}
  \put(99,-13){$u$}
  \put(99,110){$y$}
  \put(-10,110){$v$}
  \put(47,-13){$\a$}
  \put(47,105){$\a$}
  \put(-13,47){$\b$}
  \put(105,47){$\b$}
\end{picture}
\caption{An elementary quadrilateral; fileds are assigned to vertices}
\label{Fig:quadrilateral}
\end{center}
\end{figure}

Basic building blocks of systems on quad--graphs are equations on
quadrilaterals of the type
\begin{equation}\label{basic eq}
Q(x,u,v,y;\alpha,\beta)=0,
\end{equation}
where the fields $x,u,v,y\in\mathbb C$ are assigned to the four
vertices of the quadrilateral, and the parameters
$\alpha,\beta\in\mathbb C$ are assigned to its edges, as shown on
Fig.\,\ref{Fig:quadrilateral}.

A typical example is the so called ``cross-ratio equation''
\begin{equation}\label{cross ratio eq}
 \frac{(x-u)(y-v)}{(u-y)(v-x)}=\frac{\a}{\b},
\end{equation}
where on the left hand side one recognizes the cross-ratio of the
four complex points $x,u,y,v$. We shall use the cross-ratio
equation to illustrate various notions and claims in this
introduction.

Roughly speaking, the goal of the present paper is to
classify equations (\ref{basic eq}) building integrable
systems on quad-graphs. We now list more precisely the
assumptions under which we solve this problem.

First of all, we assume that the equations (\ref{basic eq}) can be
uniquely solved for any one of its arguments
$x,u,v,y\in\widehat{\mathbb{C}}$. Therefore, the solutions have to
be fractional-linear in each of their arguments. This naturally
leads to the following condition.
\vspace{6pt}

{\itbf 1) Linearity.} The function $Q(x,u,v,y;\a,\b)$ is linear in
each argument (affine linear):
\begin{equation}\label{Qlin}
  Q(x,u,v,y;\a,\b)= a_1xuvy+\dots+a_{16},
\end{equation}
where coefficients $a_i$ depend on $\a,\b$.
\vspace{6pt}

Notice that for the cross-ratio equation (\ref{cross ratio eq})
one can take the function on the left--hand side of (\ref{basic
eq}) as $Q(x,u,v,y;\a,\b)=\b(x-u)(y-v)-\a(u-y)(v-x)$.

Second, we are interested in equations on quad-graphs of arbitrary
combinatorics, hence it will be natural to assume that all variables
involved in the equations (\ref{basic eq}) are on equal footing.
Therefore, our next assumption reads as follows.
\vspace{6pt}

{\itbf 2) Symmetry.} The equation (\ref{basic eq}) is invariant
under the group $D_4$ of the square symmetries, that is function
$Q$ satisfies the symmetry properties
\begin{equation}\label{Qsym}
 Q(x,u,v,y;\a,\b)=\eps Q(x,v,u,y;\b,\a)=\sigma Q(u,x,y,v;\a,\b)
\end{equation}
with $\eps,\sigma=\pm 1$.
\vspace{6pt}
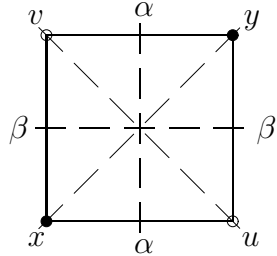
\begin{figure}[htbp]
\begin{center}
\setlength{\unitlength}{0.06em}
\begin{picture}(200,130)(-50,-20)
 \put(0,100){\circle{6}}  \put(100,100){\circle*{6}}
 \put(0,  0){\circle*{6}} \put(100,0  ){\circle{6}}
 \multiput(0,0)(100,0){2}{\line(0,1){100}}
 \multiput(0,0)(0,100){2}{\line(1,0){100}}
 \multiput(-6,50)(24,0){5}{\line(1,0){16}}
 \multiput(50,-6)(0,24){5}{\line(0,1){16}}
 \multiput(-4, -4)(19, 19){6}{\line(1, 1){13}}
 \multiput(-4,104)(19,-19){6}{\line(1,-1){13}}
 \put(-10,106){$v$}  \put(106,106){$y$}
 \put(-10,-13){$x$}  \put(105,-13){$u$}
 \put(-20,45){$\b$}  \put(113,45){$\b$}
 \put( 47,-17){$\a$} \put( 47,110){$\a$}
\end{picture}
\caption{$D_4$ symmetry}\label{fig.square}
\end{center}
\end{figure}

Of course, due to the symmetries (\ref{Qsym}) not all
coefficients $a_i$ in (\ref{Qlin}) are independent, cf.
formulae (\ref{case1}), (\ref{case2}) below.

We are interested in {\itbf integrable} equations of the type
(\ref{basic eq}), i.e. those admitting a {\itbf discrete zero
curvature representation}. We refer the reader to \cite{A2, BHS, BS},
where this notion was defined for systems on arbitrary
graphs. As pointed out above, in the third of these papers
it was shown that the integrability can be detected in
an algorithmic manner starting with no more information than the
equation itself: the criterion of integrability of an equation is
its {\itbf three-dimensional consistency}. This property means that the
equation (\ref{basic eq}) may be consistently embedded in a
three-dimensional lattice, so that similar equations hold for all
six faces of any elementary cube, as on Fig.\,\ref{fig.cube} (it
is supposed that the values of the parameters $\alpha_j$ assigned
to the opposite edges of any face are equal to one another, so
that, for instance, all edges $(x_2,x_{12})$, $(x_3,x_{31})$, and
$(x_{23},x_{123})$ carry the label $\a_1$):
\begin{figure}[htbp]
\begin{center}
\setlength{\unitlength}{0.08em}
\begin{picture}(200,170)(-50,-20)
  \put(100,  0){\circle*{6}} \put(0  ,100){\circle*{6}}
  \put( 50, 30){\circle*{6}} \put(150,130){\circle*{6}}
  \put(  0,  0){\circle{6}}  \put(100,100){\circle{6}}
  \put( 50,130){\circle{6}}  \put(150, 30){\circle{6}}
  \put( 0,  0){\line(1,0){100}}
  \put( 0,100){\line(1,0){100}}
  \put(50,130){\line(1,0){100}}
  \multiput(50,30)(20,0){5}{\line(1,0){15}}
  \put(  0, 0){\line(0,1){100}}
  \put(100, 0){\line(0,1){100}}
  \put(150,30){\line(0,1){100}}
  \multiput(50,30)(0,20){5}{\line(0,1){15}}
  \put(  0,100){\line(5,3){50}}
  \put(100,100){\line(5,3){50}}
  \put(100,  0){\line(5,3){50}}
  \multiput(50,30)(-16.67,-10){3}{\line(-5,-3){12}}
     \put(-10,-13){$x$}
     \put(90,-13){$x_1$}
     \put(50,17){$x_2$}
     \put(-13,110){$x_3$}
     \put(160,25){$x_{12}$}
     \put(45,140){$x_{23}$}
     \put(109,95){$x_{31}$}
     \put(157,135){$x_{123}$}
     \put(40,-13){$\alpha_1$}
     \put(-16,50){$\alpha_3$}
     \put(20,25){$\alpha_2$}
\end{picture}
\caption{Three-dimensional consistency}\label{fig.cube}
\end{center}
\end{figure}
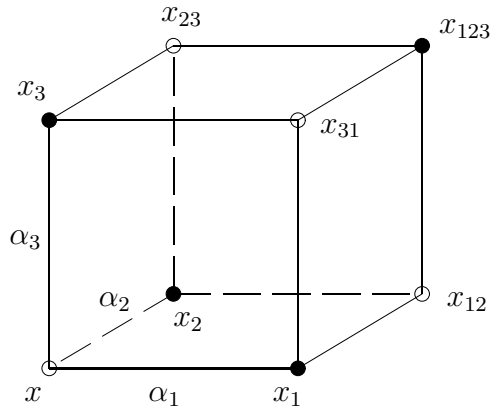

To describe more precisely what is meant under the
three-dimensional consistency, consider the Cauchy problem with
the initial data $x,x_1,x_2,x_3$. The equations
\begin{equation}\label{Qij}
  Q(x,x_i,x_j,x_{ij};\a_i,\a_j)=0
\end{equation}
allow one to determine uniquely the values $x_{12},x_{23},x_{31}$.
After that one has three different equations for $x_{123}$, coming
from the faces $(x_1,x_{12},x_{31},x_{123})$,
$(x_2,x_{23},x_{12},x_{123})$, and $(x_3,x_{31},x_{23},x_{123})$.
Consistency means that all three values thus obtained for
$x_{123}$ coincide.

For instance, consider the cross-ratio equation (\ref{cross ratio eq}).
It is not difficult to check that it possesses the property
of the three-dimensional consistency, and
\begin{equation}\label{cross ratio eq x123}
  x_{123}=\frac{(\a_1-\a_2)x_1x_2+(\a_3-\a_1)x_3x_1+(\a_2-\a_3)x_2x_3}
               {(\a_3-\a_2)x_1+(\a_1-\a_3)x_2+(\a_2-\a_1)x_3}.
\end{equation}

Looking ahead, we mention a very amazing and unexpected feature of
the expression (\ref{cross ratio eq x123}): the value $x_{123}$
actually depends on $x_1,x_2,x_3$ only, and does not depend on
$x$. In other words, four black points on Fig.\,\ref{fig.cube}
(the vertices of a tetrahedron) are related by a well-defined equation.
One can rewrite the equation (\ref{cross ratio eq x123}) as
\begin{equation}\label{cross ratio eq 123}
\frac{(x_1-x_3)(x_2-x_{123})}{(x_3-x_2)(x_{123}-x_1)}=
\frac{\a_1-\a_3}{\a_2-\a_3},
\end{equation}
which also has an appearance of the cross-ratio equation for the
four points $(x_1,x_3,x_2,x_{123})$ with the parameters
$\a_1-\a_3$ assigned to the edges $(x_1,x_3)$, $(x_2,x_{123})$ and
$\a_2-\a_3$ assigned to the edges $(x_2,x_3)$, $(x_1,x_{123})$.

This property, being very strange from the first glance, holds
actually not only in this but in all known nontrivial examples.
We take it as an additional assumption in our solution of the
classification problem.
\vspace{6pt}

{\itbf 3) Tetrahedron property.} The function
$x_{123}=z(x,x_1,x_2,x_3;\a_1,\a_2,\a_3)$, existing due to the
three-dimensional consistency, actually does not depend on the
variable $x$, that is, $z_x=0$.
\vspace{6pt}

Under the tetrahedron condition we can paint the vertices of the
cube into black and white ones, as on Fig.\,\ref{fig.cube}, and
the vertices of each of two tetrahedrons satisfy an equation of
the form
\begin{equation}\label{Q123}
  \widehat{Q}(x_1,x_2,x_3,x_{123};\a_1,\a_2,\a_3)=0;
\end{equation}
it is easy to see that under the assumption 2) (linearity) the
function $\widehat{Q}$ may be also taken linear in each argument.
(Clearly, formulas (\ref{cross ratio eq x123}), (\ref{cross
ratio eq 123}) may be also written in such a form.)

Actually, the tetrahedron condition is closely related to another
property of Eq. (\ref{basic eq}), namely to the existence of a
{\itbf three-leg form} of this equation \cite{BS}:
\begin{equation}\label{3}
  \psi(x,u;\a)-\psi(x,v;\b)=\phi(x,y;\a,\b).
\end{equation}
The three terms in this equation correspond to three ``legs'': two
short ones, $(x,u)$ and $(x,v)$, and a long one, $(x,y)$.
The short legs are the edges of the original quad-graph, while the
long one is not. (We say that the three-leg form (\ref{3}) is
centered at the vertex $x$; of course, due to the symmetries of
the function $Q$ there have to exist also similar formulas
centered at other three vertices involved.) For instance, the
cross-ratio equation (\ref{cross ratio eq}) is equivalent to
the following one:
\begin{equation}\label{cross ratio 3-leg}
  \frac{\a}{u-x}-\frac{\b}{v-x}=\frac{\a-\b}{y-x}.
\end{equation}

The three-leg form gives an explanation for the
equation for the ``black'' tetrahedron from Fig.\,\ref{fig.cube}.
Consider three faces adjacent to the vertex $x_{123}$ on
this figure, namely the quadrilaterals
$(x_1,x_{12},x_{31},x_{123})$, $(x_2,x_{23},x_{12},x_{123})$, and
$(x_3,x_{31},x_{23},x_{123})$. A summation of three-leg forms
(centered at $x_{123}$) of equations corresponding to these three
faces leads to the equation
\begin{equation*}
\phi(x_{123},x_1;\a_2,\a_3)+\phi(x_{123},x_2;\a_3,\a_1)+
\phi(x_{123},x_3;\a_1,\a_2)=0.
\end{equation*}
This equation, in any event, relates the fields in the ``black''
vertices of the cube only, i.e. has the form of (\ref{Q123}). For
example, for the cross-ratio equation this formula reads:
\begin{equation*}
   \frac{\a_2-\a_3}{x_1-x_{123}}+ \frac{\a_3-\a_1}{x_2-x_{123}}+
    \frac{\a_1-\a_2}{x_3-x_{123}}=0,
\end{equation*}
and a simple calculation convinces that this is equivalent to
(\ref{cross ratio eq x123}).

So, the tetrahedron property is a necessary condition for the
existence of a three-leg form. On the other hand, a verification
of the tetrahedron property is much more straightforward than
finding the three-leg form, since the latter contains two {\it
\'{a} priori} unknown functions $\psi$, $\phi$.

It remains to mention that, as demonstrated in \cite{BS}, the
existence of the three-leg form allows one to immediately
establish a relation to {\itbf discrete systems of the Toda type} \cite{A2}.
Indeed, if $x$ is a common vertex of $n$ adjacent quadrilaterals
faces $(x,x_{k},x_{k,k+1},x_{k+1})$, $k=1,2,\ldots,n$, with the
parameters $\a_k$ assigned to the edges $(x,x_k)$ (cf.
Fig.\,\ref{Fig:flower}), then the fields in the point $x$ and in
the ``black'' vertices of the adjacent faces satisfy the following
equation:
\begin{equation}\label{Toda}
  \sum_{k=1}^{n}\phi(x,x_{k,k+1},\a_k,\a_{k+1})=0.
\end{equation}
This is a discrete Toda type equation (equation on stars) on the graph whose
vertices are the ``black'' vertices of the original quad-graph, and whose
edges are the diagonals of the faces of the original quad-graph connecting
the ``black'' vertices. The parameters $\alpha$ are naturally assigned to the
corners of the faces of the ``black'' subgraph. See Fig.\,\ref{Fig:star}.
 (Of course, a similar Toda type
equation holds also for the ``white'' subgraph.)
\begin{figure}[htbp]
    \setlength{\unitlength}{20pt}
\hfill
\begin{minipage}[t]{170pt}
\begin{picture}(8,9)
\put(4,4){\circle*{0.2}} \put(4.2,8){\circle*{0.2}}
\put(5,6){\circle{0.2}} \put(3,5.7){\circle{0.2}}
\put(7,5.9){\circle*{0.2}} \put(6,4.2){\circle{0.2}}
\put(7.5,2.6){\circle*{0.2}} \put(5.1,2.5){\circle{0.2}}
\put(4,0.2){\circle*{0.2}} \put(3,2.1){\circle{0.2}}
\put(0.4,3.9){\circle*{0.2}} \thicklines \path(4,4)(5,6)
\path(5,6)(4.2,8) \path(4.2,8)(3,5.7) \path(3,5.7)(4,4)
\path(3,5.7)(0.4,3.9) \path(0.4,3.9)(3,2.1) \path(3,2.1)(4,4)
\path(3,2.1)(4,0.2) \path(4,0.2)(5.1,2.5) \path(5.1,2.5)(4,4)
\path(5.1,2.5)(7.5,2.6) \path(7.5,2.6)(6,4.2) \path(6,4.2)(4,4)
\path(6,4.2)(7,5.9) \path(7,5.9)(5,6) \put(3.3,4){$x$}
\put(6.3,4.2){$x_1$}  \put(7.3,5.9){$x_{12}$} \put(5.1,6.3){$x_2$}
\put(4.1,8.4){$x_{23}$}  \put(2.2,6.1){$x_3$}
\put(-0.5,4.2){$x_{34}$} \put(2.3,1.6){$x_4$}
\put(3.9,-0.3){$x_{45}$} \put(5.2,2.0){$x_5$}
\put(7.6,2.1){$x_{51}$} \put(4.8,4.3){$\alpha_1$}
\put(4.7,5){$\alpha_2$} \put(3.5,5.1){$\alpha_3$}
\put(3.5,2.7){$\alpha_4$} \put(4.7,3.2){$\alpha_5$}
\end{picture}
    \caption{Faces adjacent to the vertex $x$.}
        \label{Fig:flower}
        \end{minipage}\hspace{30pt}
    \begin{minipage}[t]{170pt}
\begin{picture}(8,9)
\put(4,4){\circle*{0.2}}
\put(4.2,8){\circle*{0.2}}
\put(7,5.9){\circle*{0.2}}
\put(7.5,2.6){\circle*{0.2}}
\put(4,0.2){\circle*{0.2}}
\put(0.4,3.9){\circle*{0.2}}
\thicklines
\path(4,4)(4.2,8) \path(4,4)(7,5.9) \path(4,4)(7.5,2.6)
\path(4,4)(4,0.2) \path(4,4)(0.4,3.9)
\put(7.3,5.9){$x_{12}$}
\put(4.1,8.3){$x_{23}$}  \put(0.2,4.2){$x_{34}$}
\put(3.9,-0.3){$x_{45}$} \put(7.5,2.2){$x_{51}$}
\put(4.5,4){$\alpha_1$}  \put(4.1,4.5){$\alpha_2$}
\put(6.7,5.3){$\alpha_2$} \put(6.1,5.9){$\alpha_1$}
\put(3.3,4.2){$\alpha_3$}
\put(3.3,3.5){$\alpha_4$}
\put(4.1,3.3){$\alpha_5$}
\end{picture}
    \caption{The star of the vertex $x$ in the black subgraph.}
        \label{Fig:star}
        \end{minipage}
\end{figure}

For instance, in the case of the cross-ratio equation, the
discrete Toda type system (\ref{Toda}) reads as
\begin{equation}\label{cross ratio Toda}
  \sum_{k=1}^{n}\frac{\a_k-\a_{k+1}}{x_{k,k+1}-x}=0.
\end{equation}

In the next Section we will show that the tetrahedron condition
naturally separates one of two subcases of the general problem of
classification of integrable equations on quad-graphs.
The second subcase will not be considered in this paper; the
corresponding subclass of equations is certainly not empty, but we
are aware only of trivial (linearizable) examples.
\vspace{6pt}

By solving the classification problem we identify equations related by
certain natural transformations. First, acting simultaneously on all
variables $x$ by one and the same M\"obius transformation does not violate
our three assumptions. Second, the same holds for the simultaneous
point change of all parameters $\a\mapsto\varphi(\a)$. Our results
on the classification of integrable equations on quad-graphs are
given by the following statement.
\begin{theorem}\label{th:list}
Up to common M\"obius transformations of the variables $x$ and
point transformations of the parameters $\a$, the
three-dimensionally consistent quad-graph equations (\ref{Qij})
with the properties 1),2),3) (linearity, symmetry, tetrahedron
property) are exhausted by the following three lists Q, H, A
($u=x_1$, $v=x_2$, $y=x_{12}$, $\a=\a_1$, $\b=\a_2$).

List $Q$:
\begin{itemize}
  \item[{\rm(Q1)}] \quad $\a(x-v)(u-y)-\b(x-u)(v-y)+\d^2\a\b(\a-\b)=0$,
  \item[{\rm(Q2)}] \quad $\begin{array}{r} \\
                   \a(x-v)(u-y)-\b(x-u)(v-y)+\a\b(\a-\b)(x+u+v+y)\\
                   -\a\b(\a-\b)(\a^2-\a\b+\b^2)=0,
                         \end{array}$
  \item[{\rm(Q3)}] \quad $\begin{array}{r} \\
                   (\b^2-\a^2)(xy+uv)+\b(\a^2-1)(xu+vy)-\a(\b^2-1)(xv+uy) \\
                   -\d^2(\a^2-\b^2)(\a^2-1)(\b^2-1)/(4\a\b)=0,
                         \end{array}$
  \item[{\rm(Q4)}] \quad $\begin{array}{r} \\
                a_0xuvy +a_1(xuv+uvy+vyx+yxu)+a_2(xy+uv)+\bar a_2(xu+vy) \\
                +\tilde a_2(xv+uy)+a_3(x+u+v+y)+a_4=0,\\
  \end{array}$
  \end{itemize}
  where the coefficients $a_i$ are expressed through
  $(\a,a)$ and $(\b,b)$ with $a^2=r(\a)$, $b^2=r(\b)$, $r(x)=4x^3-g_2x-g_3$,
  by the following formulae:
\begin{gather*}
 a_0=a+b,\quad a_1=-\b a-\a b,\quad a_2=\b^2a+\a^2b, \\
 \bar a_2=\frac{ab(a+b)}{2(\a-\b)}+\b^2a-(2\a^2-\frac{g_2}{4})b,\\
 \tilde a_2=\frac{ab(a+b)}{2(\b-\a)}+\a^2b-(2\b^2-\frac{g_2}{4})a, \\
 a_3=\frac{g_3}{2}a_0-\frac{g_2}{4}a_1,\quad
 a_4=\frac{g^2_2}{16}a_0-g_3a_1.
\end{gather*}

List $H$:
\begin{itemize}
  \item[{\rm(H1)}] \quad $(x-y)(u-v)+\b-\a=0$,
  \item[{\rm(H2)}] \quad $(x-y)(u-v)+(\b-\a)(x+u+v+y)+\b^2-\a^2=0$,
  \item[{\rm(H3)}] \quad $\a(xu+vy)-\b(xv+uy)+\d(\a^2-\b^2)=0$.
\end{itemize}

List $A$:
\begin{itemize}
  \item[{\rm(A1)}] \quad $\a(x+v)(u+y)-\b(x+u)(v+y)-\d^2\a\b(\a-\b)=0$,
  \item[{\rm(A2)}] \quad $(\b^2-\a^2)(xuvy+1)+\b(\a^2-1)(xv+uy)
    -\a(\b^2-1)(xu+vy)=0$.
\end{itemize}
\end{theorem}

\paragraph{Remarks}
1) The list A can be dropped down by allowing an extended
group of M\"obius transformations, which act on the variables
$x,y$ differently than on $u,v$ (white and black sublattices on
Figs.\,\ref{Fig:quadrilateral},\ref{fig.cube}). In this manner Eq.
(A1) is related in to (Q1) (by the change $u\to -u$, $v\to -v$),
and Eq. (A2) is related to (Q3) with $\delta=0$ (by the change
$u\to 1/u$, $v\to 1/v$). So, really independent equations are
given by the lists Q and H.

2) In both lists Q, H the last equations are the most general ones.
This means that Eqs. (Q1)--(Q3) and (H1), (H2) may be obtained from (Q4)
and (H3), respectively, by certain degenerations and/or limit procedures.
So, one could be tempted to shorten down these lists to one item each.
However, on the one hand, these limit procedures are outside of our
group of admissible (M\"obius) transformations, and on the other hand,
in many situations the ``degenerate'' equations (Q1)--(Q3) and (H1), (H2)
are of an interest for themselves. This resembles the situation with
the list of six Painlev\'e equations and the coalescences connecting
them, cf. \cite{IKSY}.

3) Parameter $\delta$ in Eqs. (Q1), (Q3), (H3) can be scaled away,
so that one can assume without loss of generality that $\delta=0$
or $\delta=1$.

4) It is natural to set in Eq. (Q4) $(\a,a)=(\wp(A),\wp'(A))$ and,
similarly, $(\b,b)=(\wp(B),\wp'(B))$. So, this equation is
actually parametrized by two points of the elliptic curve
$\mu^2=r(\lambda)$. The appearance of an elliptic curve in our
classification problem is by no means obvious from the beginning,
its origin will become clear later, in the course of the proof.
For the cases of $r$ with multiple roots, when the elliptic curve
degenerates into a rational one, Eq. (Q4) degenerates to one of
the previous equations of the list Q; for example, if $g_2=g_3=0$
then the inversion $x\to 1/x$ turns (Q4) into (Q2).

\paragraph{Bibliographical remarks.} It is difficult to track down the
origin of the equations listed in Theorem 1. Probably, the oldest ones are
(H1) and ${\rm (H3)}_{\d=0}$, which can be found in the work of Hirota
\cite{H}  (of course not on general quad--graphs but only on the standard
square lattice with the labels $\a$ constant in each of the two lattice
directions; similar remarks apply also to other references in this paragraph).
Eqs. (Q1) and ${\rm (Q3)}_{\d=0}$ go back to \cite{QNCL},
see also a review in \cite{NC}. Eq. (Q4) was found in \cite{A1}.
A Lax representation for (Q4) was found in \cite{N} with the help of the
method based on the three--dimensional consistency, identical with the
method introduced in \cite{BS}.
Eqs. (Q2) and ${\rm (Q3)}_{\d=1}$ are particular cases of (Q4), but seem to
have not appeared explicitly in the literature. The same holds for
(H2) and ${\rm (H3)}_{\d=1}$.

\section{Classification: analysis}\label{sect:solv1}

In principle, the three-dimensional consistency turns, under the
assumptions 1), 2), into some system of functional equations for
the coefficients $a_i$ of the functions $Q$ (see (\ref{Qlin})).
However, this system is difficult to analyze and we will take
a different route.

For the first step, we consider the problem of the three-dimensional
consistency in the following general setting: find triples of functions
$f_1,f_2,f_3$ of three arguments such that if
\begin{equation}\label{xij}
  x_{23}=f_1(x,x_2,x_3), \quad
  x_{31}=f_2(x,x_3,x_1), \quad
  x_{12}=f_3(x,x_1,x_2)
\end{equation}
then
\begin{equation}\label{x123}
  x_{123}:= f_1(x_1,x_{12},x_{31}) \equiv
            f_2(x_2,x_{23},x_{12}) \equiv
            f_3(x_3,x_{31},x_{23})
\end{equation}
identically in $x,x_1,x_2,x_3$. In other words, we ignore for a moment
the conditions 1) and 2) and look to what consequences the {\it
tetrahedron} condition leads. The proof of the following statement
demonstrates that this condition separates just one of two
possible subcases in the general problem.

\begin{proposition}\label{st:fff}
For the functions $f_1,f_2,f_3$ compatible in the sense
(\ref{xij}), (\ref{x123}) and satisfying the tetrahedron
condition, the following relation holds:
\begin{equation}\label{fff}
 f_{3,x_2}f_{2,x_1}f_{1,x_3}=-f_{3,x_1}f_{2,x_3}f_{1,x_2}
\end{equation}
identically in $x,x_1,x_2,x_3$.
\end{proposition}
\begin{proof}
Denote $x_{123}=z(x,x_1,x_2,x_3)$, and also denote for short the
functions with shifted arguments (\ref{x123}) by capitals:
$F_1=f_1(x_1,x_{12},x_{31})$, etc. Then differentiating
(\ref{x123}) with respect to $x_2,x_3$ and $x$ yields the
following system which is linear with respect to the derivatives
of $F_1$:
\[
 \begin{array}{lll}
   f_{3,x_2}F_{1,x_{12}} &                          &=z_{x_2}\\
                         &\ \ f_{2,x_3}F_{1,x_{31}} &=z_{x_3}\\
   f_{3,x}F_{1,x_{12}}   &+   f_{2,x}F_{1,x_{31}}   &=z_x
 \end{array}
\]
and two analogous systems for $F_2,F_3$ obtained by the cyclic
shift of indices. The above system is overdetermined,
and, excluding the derivatives of $F_1$, we come to the first
equation of the following system (the other two come from the
similar considerations with $F_2,F_3$):
\[
 \begin{array}{rrrr}
                          &  f_{2,x_3}f_{3,x}z_{x_2} & +f_{3,x_2}f_{2,x}z_{x_3}
& =f_{3,x_2}f_{2,x_3}z_x\\
  f_{1,x_3}f_{3,x}z_{x_1} &                          & +f_{3,x_1}f_{1,x}z_{x_3}
& =f_{1,x_3}f_{3,x_1}z_x\\
  f_{1,x_2}f_{2,x}z_{x_1} & +f_{2,x_1}f_{1,x}z_{x_2} &
& =f_{2,x_1}f_{1,x_2}z_x\\
 \end{array}
\]
Now, the tetrahedron condition $z_x=0$ implies that the r.h.s.
vanishes, and therefore the determinant has to vanish
as well; this is equivalent to (\ref{fff}).
\end{proof}

The second possibility, which we do not consider here, would be
that $z_x\ne 0$ and Eq. (\ref{fff}) does not hold. In this case
the above system can be solved to give $z_{x_i}=\Phi_iz_x$, $i=1,2,3$, where
$\Phi_i$ are expressed through $f_{j,x}$, $f_{j,x_k}$. The compatibility of
the latter three equations can be expressed as a system of differential
equations for $\Phi_i$, and therefore for $f_j$, which are much more
complicated than (\ref{fff}). It certainly deserves a further investigation,
see discussion in Section \ref{sect:conclusions}.

The necessary condition (\ref{fff}) will provide us with
a finite list of candidates for the three-dimensional consistency.
First of all, we rewrite the relation (\ref{fff}) in terms of the
polynomial $Q$ using both the {\it linearity} and  the {\it symmetry}
assumptions.

\begin{proposition}\label{st:ggg}
Relation (\ref{fff}) is equivalent to
\begin{multline}\label{ggg}
     g(x,x_1;\a_1,\a_2)g(x,x_2;\a_2,\a_3)g(x,x_3;\a_3,\a_1) \\
  = -g(x,x_1;\a_1,\a_3)g(x,x_2;\a_2,\a_1)g(x,x_3;\a_3,\a_2),
\end{multline}
where $g(x,u;\a,\b)$ is a biquadratic polynomial in $x,u$ defined
by either of the formulas
\begin{eqnarray}
   g(x,u;\a,\b) & = & QQ_{yv}-Q_yQ_v, \label{gxu}\\
   g(x,v;\b,\a) & = & QQ_{yu}-Q_yQ_u, \label{gxv}
\end{eqnarray}
where $Q=Q(x,u,v,y;\a,\b)$. The polynomial $g$ is symmetric:
\begin{equation}\label{gux}
   g(x,u;\a,\b)=g(u,x;\a,\b).
\end{equation}
\end{proposition}
\begin{proof}
The equivalence of the definitions (\ref{gxu}), (\ref{gxv})
follows from the first symmetry property in (\ref{Qsym})
($\a\leftrightarrow\b,$ $u\leftrightarrow v$), while the second
one ($x\leftrightarrow u,$ $y\leftrightarrow v$) implies
(\ref{gux}).

To prove (\ref{ggg}), let $Q=p(x,u,v)y+q(x,u,v)$, so that
$y=f(x,u,v)=-q/p.$ Then $f_v/f_u=(q_vp-qp_v)/(q_up-qp_u)$, and
substituting $p=Q_y$, $q=Q-yQ_y$, we obtain
\[
  \frac{f_v}{f_u}=\frac{QQ_{yv}-Q_yQ_v}{QQ_{yu}-Q_yQ_u},
\]
which yields:
\[
  \frac{f_{k,x_j}}{f_{k,x_i}}=\frac{g(\a_i,\a_j;x,x_i)}{g(\a_j,\a_i;x,x_j)},
  \quad (i,j,k)=\text{c.p.}(1,2,3).
\]
Now (\ref{ggg}) follows from (\ref{fff}).
\end{proof}

The biquadratic polynomials (\ref{gxu}) and (\ref{gxv}) are
associated to the edges of the basic square. One can consider also
the polynomial
\begin{equation}\label{Gxy}
   G(x,y;\a,\b)=QQ_{uv}-Q_uQ_v,
\end{equation}
associated to the diagonal. They have the following important property.
\begin{lemma}\label{discriminants}
The discriminants of the polynomials $g=g(x,u;\a,\b)$,
$\bar{g}=g(x,v;\b,\a)$, and $G=G(x,y;\a,\b)$, considered as
quadratic polynomials in $u$, $v$, and $y$, respectively,
coincide:
\begin{equation}\label{dis3}
  g^2_u-2gg_{uu}= \bar g^2_v-2\bar g\bar g_{vv}= G^2_y-2GG_{yy}.
\end{equation}
\end{lemma}
\begin{proof} This follows solely from the fact that the function
$Q$ is linear in each argument. Indeed, calculate
\[
g^2_u-2gg_{uu}=((QQ_{yv}-Q_yQ_v)_u)^2-2(QQ_{yv}-Q_yQ_v)(QQ_{yv}-Q_yQ_v)_{uu},
\]
taking into account that $Q_{uu}=0$. The result reads:
\begin{eqnarray*}
g^2_u-2gg_{uu}
     & = &
       Q^2Q^2_{uvy}+Q^2_uQ^2_{vy}+Q^2_vQ^2_{uy}+Q^2_yQ^2_{uv}+
       4QQ_{uv}Q_{uy}Q_{vy}\\
     &   & -2QQ_{uvy}(Q_uQ_{vy}+Q_vQ_{uy}+Q_yQ_{uv})-4Q_uQ_vQ_yQ_{uvy}\\
     &   & -2Q_uQ_vQ_{uy}Q_{vy}-2Q_uQ_yQ_{uv}Q_{vy}-2Q_vQ_yQ_{uv}Q_{uy}.
\end{eqnarray*}
It remains to notice that this expression is symmetric with
respect to all indices.
\end{proof}

In the formula (\ref{ggg}) the variables are highly separated, and it can
be effectively analyzed further on. In the next statement we demonstrate
that this functional equation relating values of $g$ with different arguments
implies some properties for a {\em single} polynomial $g$.

\begin{proposition}\label{st:ghk}
The polynomial $g(x,u;\a,\b)$ can be represented as
\begin{equation}\label{ghk}
   g(x,u;\a,\b)=k(\a,\b)h(x,u;\a),
\end{equation}
where the factor $k$ is antisymmetric,
\begin{equation}\label{kk}
   k(\b,\a)=-k(\a,\b),
\end{equation}
and the coefficients of the polynomial $h(x,u;\a)$ depend on
parameter $\a$ in such a way that its discriminant
\begin{equation}\label{r}
   r(x)=h^2_u-2hh_{uu}
\end{equation}
does not depend on $\a$.
\end{proposition}
\begin{proof}
Relation (\ref{ggg}) implies that the fraction
$g(x,x_1;\a_1,\a_2)/g(x,x_1;\a_1,\a_3)$ does not depend on $x_1$,
and due to the symmetry (\ref{gux}) it does not depend on $x$ as
well. Therefore
\[
  \frac{g(x,x_1;\a_1,\a_2)}{g(x,x_1;\a_1,\a_3)}
  =\frac{\kappa(\a_1,\a_2)}{\kappa(\a_1,\a_3)},
\]
where, because of (\ref{ggg}), the function $\kappa$ satisfies the
equation
\[
   \kappa(\a_1,\a_2)\kappa(\a_2,\a_3)\kappa(\a_3,\a_1)
 =-\kappa(\a_2,\a_1)\kappa(\a_3,\a_2)\kappa(\a_1,\a_3).
\]
This equation is equivalent to
\[
 \kappa(\b,\a)= -\frac{\phi(\a)}{\phi(\b)}\kappa(\a,\b),
\]
that is, the function $k(\a,\b)=\phi(\a)\kappa(\a,\b)$ is
antisymmetric. We have:
\[
\frac{g(x,u;\a,\b)}{\kappa(\a,\b)}=
\frac{g(x,u;\a,\g)}{\kappa(\a,\g)}\quad \Rightarrow\quad
  \frac{g(x,u;\a,\b)}{k(\a,\b)}= \frac{g(x,u;\a,\g)}{k(\a,\g)},
\]
which implies (\ref{ghk}). To prove the last statement of the
proposition, we notice that, according to (\ref{ghk}), (\ref{kk}),
\[
  k(\a,\b)h(x,u;\a)= g(x,u;\a,\b), \quad
 -k(\a,\b)h(x,v;\b)= g(x,v;\b,\a).
\]
Due to the identity (\ref{dis3}), we find:
\[
  h^2_u-2hh_{uu}=\bar h^2_v-2\bar h\bar h_{vv},
  \quad h=h(x,u;\a),\quad \bar h=h(x,v;\b),
\]
and therefore $r$ does not depend on $\a$.
\end{proof}
\vspace{1mm}

Thus, the three-dimensional consistency with the tetrahedron
property implies the following remarkable property of the function
$Q$ which will be called property (R):
\begin{itemize}
  \item[{\rm (R)}]
the determinant $g=QQ_{vy}-Q_vQ_y$ is factorizable as in
(\ref{ghk}), (\ref{kk}), and the discriminant $r=h^2_u-2hh_{uu}$
of the corresponding quadratic polynomial $h$ does not depend on
parameters at all.
\end{itemize}
It remains to classify all functions $Q$ with this property.
This will be done in the next
section. A finite list of functions $Q$ with the property (R)
consists, therefore, of {\em candidates} for the three-dimensional
consistency. The final check is straightforward, and shows that
the property (R) is not only necessary but also almost sufficient
for the three-dimensional consistency with the tetrahedron
property (the list of functions with the property (R) consists of a dozen
of items; only in two of them one finds functions violating the
consistency).
\vspace{1mm}

As a preliminary step, we consider more closely the coefficients
of the polynomial $Q$. The symmetry property (\ref{Qsym}) easily
implies that two cases are possible, with $\sigma=1$ and
$\sigma=-1$, respectively:
\begin{align}
\nonumber
  Q=&\ a_0xuvy +a_1(xuv+uvy+vyx+yxu) \\
\nonumber
    & +a_2(xy+uv)+\bar a_2(xu+vy)+\tilde a_2(xv+uy) \\
\label{case1}
    & +a_3(x+u+v+y)+a_4, \\
\nonumber
  Q=&\ a_1(xuv+uvy-vyx-yxu) \\
\label{case2}
    & +a_2(xy-uv)+a_3(x-u-v+y),
\end{align}
where
\begin{equation}\label{asym}
 a_i(\b,\a)=\eps a_i(\a,\b), \quad
 \tilde a_2(\a,\b)=\eps\bar a_2(\b,\a), \quad \eps=\pm 1.
\end{equation}

It is easy to prove that the case (\ref{case2}) is actually empty.
Indeed, a light calculation shows that in this case
\[
  g(x,u;\a,\b)=g(x,u;\b,\a)=-P(x;\a,\b)P(u;\a,\b),
\]
where $P(\a,\b;x)=a_1x^2-a_2x-a_3$, so that the relation
(\ref{ggg}) becomes
\begin{multline*}
    P(x_1;\a_1,\a_2)P(x_2;\a_2,\a_3)P(x_3;\a_3,\a_1) \\
 = -P(x_1;\a_3,\a_1)P(x_2;\a_1,\a_2)P(x_3;\a_2,\a_3).
\end{multline*}
Equating to zero the coefficients at $(x_1x_2x_3)^2$, $x_1x_2x_3$,
and the free term yields $a_1=a_2=a_3=0$ -- a contradiction.

Turning to the case (\ref{case1}), we denote the coefficients of the
polynomials $h,r$ as follows:
\begin{equation}\label{b}
 h(x,u;\a)=b_0x^2u^2+b_1xu(x+u)+b_2(x^2+u^2)+\hat b_2xu+b_3(x+u)+b_4,
\end{equation}
\begin{equation} \label{c}
  r(x)=c_0x^4+c_1x^3+c_2x^2+c_3x+c_4,
\end{equation}
where $b_i=b_i(\a)$. So, we consequently descended from the
polynomial $Q$ (4 variables, 7 coefficients depending on 2
parameters) to the polynomial $h$ (2 variables, 6 coefficients
depending on 1 parameter; also take into account the factor $k$),
and then to the polynomial $r$ (1 variable, 5 coefficients, no
parameters).

It remains to go the way back, i.e. to reconstruct $k,h$ and
$Q$ from a given polynomial $r$ (which does not depend on the
parameter $\alpha$). This will be done in the next section.

\section{Classification: synthesis}\label{sect:solv2}

First of all, we factor out the action of the simultaneous M\"obius
transformations of the variables $x$. The action $x\mapsto (ax+b)/(cx+d)$
transforms the polynomials $h$, $r$ as follows:
\begin{eqnarray*}
 & h(x,u;\a)\mapsto
(cx+d)^2(cu+d)^2h\left(\dfrac{ax+b}{cx+d},\dfrac{au+b}{cu+d};\a\right), & \\
& r(x)\mapsto (cx+d)^4r\left(\dfrac{ax+b}{cx+d}\right). &
\end{eqnarray*}
Using such transformations one can bring the polynomial $r$ into one of the
following canonical forms, depending on the distribution of its roots:
\begin{itemize}
\item $r(x)=0$;
\item $r(x)=1\;\;$ ($r$ has one quadruple zero);
\item $r(x)=x\;\;$ ($r$ has one simple zero and one triple zero);
\item $r(x)=x^2\;\;$ ($r$ has two double zeroes);
\item $r(x)=x^2-1\;\;$ ($r$ has two simple zeroes and one double zero);
\item $r(x)=4x^3-g_2x-g_3,\;$ $\Delta=g_2^3-27g_3^2\neq 0\;\;$ ($r$ has
four simple zeroes).
\end{itemize}
Next, we find for these canonical polynomials $r$ all
possible polynomials $h$.

\begin{proposition}\label{st:rh}
For a given polynomial $r(x)$ of the fourth degree, in one of the canonical
forms above, the symmetric biquadratic polynomials $h(x,u)$ having $r(x)$ as
their discriminants, $r(x)=h^2_u-2hh_{uu}$, are exhausted by the following list:
\begin{align}
\label{q0}
 & r=0: && h=\frac{1}{\a}(x-u)^2; \tag{q0}\\
\label{h1}
 &      && h=(\g_0xu+\g_1(x+u)+\g_2)^2; \tag{h1}\\
 \label{q1}
 & r=1: && h=\frac{1}{2\a}(x-u)^2-\frac{\a}{2}; \tag{q1}\\
 \label{h2}
 &      && h=\g_0(x+u)^2+\g_1(x+u)+\g_2,\quad \g^2_1-4\g_0\g_2=1; \tag{h2}\\
\label{q2}
 & r=x: && h=\frac{1}{4\a}(x-u)^2-\frac{\a}{2}(x+u)+\frac{\a^3}{4}; \tag{q2} \\
\label{h3}
 & r=x^2: && h=\g_0x^2u^2+\g_1xu+\g_2,\quad \g^2_1-4\g_0\g_2=1;\tag{h3} \\
\label{q3}
 & r=x^2-\d^2: && h= \frac{\a}{1-\a^2}(x^2+u^2)
                  -\frac{1+\a^2}{1-\a^2}xu +\d^2\frac{1-\a^2}{4\a}; \tag{q3} \\
\nonumber
 & r=4x^3-g_2x-g_3: \mspace{-70mu}&& \mspace{70mu}
  h=\frac{1}{\sqrt{r(\a)}}\Big[\big(xu+\a(x+u)+g_2/4\big)^2 \\
\label{q4}
 &&& \mspace{100mu} -(x+u+\a)(4\a xu-g_3)\Big].\tag{q4}
\end{align}
\end{proposition}
\begin{proof}
We have to solve the system of the form
\[
 \begin{array}{ll}
  b^2_1-4b_0b_2                     &= c_0, \\
  2b_1(\hat b_2-2b_2)-4b_0b_3       &= c_1, \\
  \hat b^2_2-4b^2_2-2b_1b_3-4b_0b_4 &= c_2, \\
  2b_3(\hat b_2-2b_2)-4b_1b_4       &= c_3, \\
  b^2_3-4b_2b_4                     &= c_4,
 \end{array}
\]
where $b_k$ are the coefficients of $h(x,u)$ and $c_k$ are the
coefficients of $r(x)$ (see (\ref{b}), (\ref{c})). This is done by a
straightforward analysis. For example, consider in detail the case
$r=4x^3-g_2x-g_3$. We have: $c_0=0$,
$c_1=4$, hence $b_0\ne 0$. Set $b_0=\rho^{-1}$, $b_1=-2\a\rho^{-1}$,
then $b_2=\a^2\rho^{-1}$. Next, use the second and the third equations
to eliminate $b_3$ and $b_4$, then the last two equations give an
expression for $\hat b_2$ and the constraint $\rho^2=r(\a)$.
\end{proof}

\paragraph{Remarks.} 1) Notice that for all six canonical forms of $r$
we have a one-parameter family of $h$'s, denoted in the list of
Proposition \ref{st:rh} by (q0)--(q4) (the one-parameter family
for $r=x^2$ is not explicitly written down since it coincides with
(q3) at $\d=0$). It will turn out that the polynomials $Q$
reconstructed from these $h$'s belong to the list Q (so that $h$
in (q$k$) corresponds to $Q$ in (Q$k$); $h$ from (q0) corresponds
to ${\rm (Q1)}_{\d=0}$). For the polynomials $r=0,1,x^2$ we have
additional two- or three-parameter families of $h$'s denoted in
the list of Proposition \ref{st:rh} by (h1)--(h3). They will lead
to the correspondent $Q$'s in (H1)--(H3), as well as to (A1),
(A2).

2) The expression $\sqrt{r(\a)}$ in Eq. (q4) clearly shows that
the elliptic curve $\mu^2=r(\lambda)$ comes into play at this point.
One can uniformize $\sqrt{r(\a)}=\wp'(A)$, $\a=\wp(A)$, where $A$ is a
point of the elliptic curve, and $\wp$ is the Weierstrass elliptic function.
Actually, the polynomial $h(x,u;\a)$ is
well--known in the theory of elliptic functions and represents the addition
theorem for $\wp$--function. Namely, the equality $h(\wp(X),\wp(U);\wp(A))=0$
is equivalent to $A=\pm X\pm U$ modulo the lattice of periods of the function
$\wp$. To resume: symmetric biquadratic polynomials $h(x,u)$ with the
discriminant $h_u^2-2hh_{uu}=r(x)$ are parametrized (in the non-degenerate
case) by a point of the elliptic curve $\mu^2=r(\lambda)$. This is the origin
of the elliptic curve in the parametrization of Eq. (Q4). As a consequence,
the spectral parameter in the discrete zero curvature
representation of the latter equation also lives on the elliptic curve.
This may be considered also as the ultimate reason for the spectral parameter
of the Krichever--Novikov equation to live on an elliptic curve
(\cite{KN}, see also Sect.\,\ref{sect:Backlund}).
\vspace{6pt}

It remains to reconstruct polynomials $Q$ for all $h$'s from
Proposition \ref{st:rh}.
\vspace{6pt}

{\bf Proof of Theorem 1.} For the polynomial (\ref{case1}) we have:
\begin{align*}
 g(x,u;\a,\b)=&\ (\bar a_2a_0-a^2_1)x^2u^2
                +(a_1(\bar a_2-\tilde a_2)+a_0a_3-a_1a_2)xu(x+u) \\
        & +(a_1a_3-a_2\tilde a_2)(x^2+u^2)
          +(\bar a^2_2-\tilde a^2_2+a_0a_4-a^2_2)xu \\
        & +(a_3(\bar a_2-\tilde a_2)+a_1a_4-a_2a_3)(x+u) +\bar a_2a_4-a^2_3,
\end{align*}
and an analogous expression for $g(x,u;\b,\a)$ is obtained by the
replacement $\bar a_2\leftrightarrow\tilde a_2$. Using (\ref{ghk})
and denoting $b_i=b_i(\a)$, $b'_i=b_i(\b)$, $k=k(\a,\b)$, we come to the
following system for the unknown quantities $a_k$:
\[
 \begin{array}{ll}
  \bar a_2a_0-a^2_1                      &= kb_0, \\
  a_1(\bar a_2-\tilde a_2)+a_0a_3-a_1a_2 &= kb_1, \\
  a_1a_3-a_2\tilde a_2                   &= kb_2 \\
  \bar a^2_2-\tilde a^2_2+a_0a_4-a^2_2   &= k\hat b_2, \\
  a_3(\bar a_2-\tilde a_2)+a_1a_4-a_2a_3 &= kb_3, \\
  \bar a_2a_4-a^2_3                      &= kb_4,
 \end{array}\quad\;
 \begin{array}{ll}
  \tilde a_2a_0-a^2_1                    &= -kb'_0, \\
  a_1(\tilde a_2-\bar a_2)+a_0a_3-a_1a_2 &= -kb'_1, \\
  a_1a_3-a_2\bar a_2                     &= -kb'_2, \\
  \tilde a^2_2-\bar a^2_2+a_0a_4-a^2_2   &= -k\hat b'_2, \\
  a_3(\tilde a_2-\bar a_2)+a_1a_4-a_2a_3 &= -kb'_3, \\
  \tilde a_2a_4-a^2_3                    &= -kb'_4.
 \end{array}
\]
Of course, also the quantity $k$ is still unknown here.
Since we are looking for the function $Q$ up to arbitrary factor,
it is convenient to denote
\begin{equation}\label{a to A}
 a=\bar a_2-\tilde a_2,\quad
 A_i=\frac{a_i}{a},\quad
 \widehat A_2= \frac{\bar a_2}{a}-\frac{1}{2}=\frac{\tilde a_2}{a}+\frac{1}{2},
 \quad K=\frac{k}{a^2}
\end{equation}
(it is easy to see that $a\not\equiv 0$ since otherwise
$h\equiv 0$). These functions are skew-symmetric:
\[
   A_i(\b,\a)=-A_i(\a,\b), \quad
   \widehat A_2(\b,\a)=-\widehat A_2(\a,\b), \quad
   K(\b,\a)=-K(\a,\b),
\]
and the above system can be rewritten, after some elementary
transformations, as follows:
\renewcommand{\arraystretch}{1.5}
\begin{gather*}
 \begin{array}{ll}
  K[(\hat b_2+\hat b'_2)(b_0+b'_0)-(b_1+b'_1)^2]           &
                                     = 2(b_0-b'_0), \\
  K[(b_0+b'_0)(b_3+b'_3)-(b_1+b'_1)(b_2+b'_2)]             &
                                     = b_1-b'_1, \\
  K[(b_1+b'_1)(b_3+b'_3)-(b_2+b'_2)(\hat b_2+\hat b'_2)]   &
                                     = 2(b_2-b'_2), \\
  K[(b_0+b'_0)(b_4+b'_4)-(b_2+b'_2)^2]                     &
                                     =\dfrac{1}{2}(\hat b_2-\hat b'_2), \\
  K[(b_1+b'_1)(b_4+b'_4)-(b_2+b'_2)(b_3+b'_3)]             &
                                     = b_3-b'_3, \\
  K[(\hat b_2+\hat b'_2)(b_4+b'_4)-(b_3+b'_3)^2]           &
                                     = 2(b_4-b'_4),
 \end{array}\\
 A_0=K(b_0+b'_0), \quad 2A_1=K(b_1+b'_1), \quad A_2=K(b_2+b'_2), \\
 4\widehat A_2=K(\hat b_2+\hat b'_2), \quad
 2A_3=K(b_3+b'_3), \quad A_4=K(b_4+b'_4).
\end{gather*}
The first six lines here form a system of functional equations for
$b_i$ and $K$ (call it $(K,b)$--system), while the last two line
split away, and should be considered just as definitions of $A_i$.
So, to any solution of the $(K,b)$--system there corresponds a
function $Q$ whose coefficients are given by the last two lines of
the system above, and the formulas
\[
\bar{A}_2=\widehat{A}_2+1/2, \quad \widetilde{A}_2=\widehat{A}_2-1/2,
\]
which follow from (\ref{a to A}). By construction, this function has the
property (R) and is, therefore, a candidate for the three-dimensional
consistency with the tetrahedron property.

Consider first the cases (q0), (q1), (q2), (q3), (q4), i.e. when
we have a one-parameter family of polynomials $h$. A
straightforward, although tedious, check proves that in these
cases all six equations of the $(K,b)$--system lead to one and the
same function $K$, provided the functions $b_i=b_i(\a)$ are
defined as in (q0)--(q4). Calculating the corresponding
coefficients $A_i$, we come to the functions $Q$ given in Theorem
1 by the formulas ${\rm (Q1)}_{\d=0}$, ${\rm (Q1)}_{\d=1}$, (Q2),
(Q3), and (Q4), respectively. A further straightforward check
convinces that all these functions indeed pass the
three-dimensional consistency test with the tetrahedron propery.

It remains to consider the cases (h1), (h2), (h3). A thorough analysis of the
$(K,b)$--system shows that in these cases we have the following solutions.

In the case (h1):
\begin{itemize}
\item either $h$ does not depend on parameters at all, and then it has to be
of the form $h=\big((\eps_0x+\eps_1)(\eps_0u+\eps_1)\big)^2$, and $K$
is arbitrary; performing a suitable M\"obius transformation, we can achieve
that $h\equiv 1$; the outcome of this subcase is the following
function $Q$ which is a candidate for the three-dimensional consistency:
\begin{equation}\label{H1 candidate}
Q=(x-y)(u-v)+k(\a,\b), \tag{${\rm\widehat{H1}}$}
\end{equation}
where $k$ is an arbitrary skew-symmetric function, $k(\a,\b)=-k(\b,\a)$;
\item or $h=(1/\a)(\eps_0xu+\eps_1(x+u)+\eps_2)^2$ with arbitrary constants
$\eps_i$. In this case we can use M\"obius transformations to
achieve $h=(1/\a)(x+u)^2$, and the correspondent function $Q$
coincides with ${\rm (A1)}_{\d=0}$ from Theorem 1. This function
passes the three-dimensional consistency test with the tetrahedron
property.
\end{itemize}

In the case (h2):
\begin{itemize}
\item either $h=x+u+\a$, and the correspondent function $Q$ coincides
with (H2) from Theorem 1;
\item or $h=(1/2\a)(x+u)^2-(\a/2)$, and the correspondent function (Q)
is given in ${\rm (A1)}_{\d=1}$ of Theorem 1.
\end{itemize}
Both functions are three-dimensionally consistent with the tetrahedron
property.

Finally, in the case (h3):
\begin{itemize}
\item either $h$ does not depend on parameters at all, then it has to be
equal to $h=xu$, and $K$ is arbitrary; we have in this subcase the following
function $Q$ which is a candidate for the three-dimensional consistency:
\begin{equation}\label{H3 candidate}
Q=\dfrac{1+k}{2}(xu+vy)-\dfrac{1-k}{2}(xv+uy),
\tag{${\rm\widehat{H3}}_0$}
\end{equation}
where $k$ is an arbitrary skew-symmetric function, $k(\a,\b)=-k(\b,\a)$;
\item or $h=xu+\a$ (possibly, upon application of the inversion $x\mapsto 1/x$,
$u\mapsto 1/u$); the correspondent function $Q$ is ${\rm
(H3)}_{\d=1}$ of Theorem 1;
\item or, finally, $h=\dfrac{\a}{1-\a^2}(x^2u^2+1)-\dfrac{1+\a^2}{1-\a^2}xu$;
the correspondent function $Q$ is given in (A2) of Theorem 1.
\end{itemize}
In the two last subcases the three-dimensional consistency condition is
fulfilled with the tetrahedron property.

To finish the proof of Theorem 1, we have to consider the
functions $Q$ given by the formulas (${\rm\widehat{H1}}$) and
(${\rm\widehat{H3}}_0$). They depend on an arbitrary
skew--symmetric function $k$, and have the property (R) for any
choice of the latter. As it turns out, these are the only
situations when the property (R) does not automatically imply the
three-dimensional consistency.

A direct calculation shows that the function (${\rm\widehat{H1}}$)
gives a map which is three-dimensionally consistent, with
\begin{equation}
x_{123}=\frac{k(\a_1,\a_2)x_1x_2+k(\a_2,\a_3)x_2x_3+k(\a_3,\a_1)x_3x_1}
{k(\a_3,\a_2)x_1+k(\a_1,\a_3)x_2+k(\a_2,\a_1)x_3},
\end{equation}
if and only if
\begin{equation}
k(\a_1,\a_2)+k(\a_2,\a_3)+k(\a_3,\a_1)=0.
\end{equation}
To solve this functional equation, differentiate it with respect to $\a_1$ and
$\a_2$:
\[
k_{\a_1\a_2}(\a_1,\a_2)=0,
\]
which together with the skew--symmetry of $k$ yields
$k(\a_1,\a_2)=f(\a_2)-f(\a_1)$. A point transformation of the parameter
$f(\a)\mapsto \a$ allows us to take simply $k(\a,\b)=\b-\a$. Thus, we arrive at
the case (H1) of Theorem 1.

Finally, consider the formula (${\rm\widehat{H3}}_0$). Denote, for
brevity,
\[
\ell(\a,\b)=\frac{1+k(\a,\b)}{1-k(\a,\b)}, \quad {\rm so\;\;that}\quad
\ell(\b,\a)=1/\ell(\a,\b).
\]
We will write also $\ell_{ij}$ for $\ell(\a_i,\a_j)$. A
straightforward inspection shows that the function
(${\rm\widehat{H3}}_0$) gives a three-dimensionally consistent map
with
\begin{equation}
x_{123}=\frac{(\ell_{21}-\ell_{12})x_1x_2+
(\ell_{32}-\ell_{23})x_2x_3+(\ell_{13}-\ell_{31})x_3x_1}
{(\ell_{23}-\ell_{32})x_1+(\ell_{31}-\ell_{13})x_2+(\ell_{12}-\ell_{21})x_3},
\end{equation}
if and only if
\begin{equation}
\ell(\a_1,\a_2)\ell(\a_2,\a_3)\ell(\a_3,\a_1)=1.
\end{equation}
Just as above, up to a point transformation of the parameter, the
solution of this functional equation is given by
$\ell(\a,\b)=\a/\b$, which leads to the equation ${\rm
(H3)}_{\d=0}$ of Theorem 1.

The proof of Theorem 1 is now complete. \qed

\section{Three-leg forms}\label{sect:3legs}

Theorem 1 classifies the three-dimensionally consistent equations under
the tetrahedron condition. In Section \ref{sect:intro} we have seen that
the latter condition is necessary for the existence of a {\it three-leg
form}
\begin{equation}\label{3a}
\psi(x,u;\a)-\psi(x,v;\b)=\phi(x,y;\a,\b)
\end{equation}
of Eq. (\ref{basic eq}). Recall that the formula (\ref{3a}) implies the
validity of equations on stars (\ref{Toda}), or
equations of the discrete Toda type, for the fields $x$ at the vertices of
the ``black'' sublattice; the same holds for the ``white'' sublattice.
We prove now that a three--leg form indeed exists for all equations listed
in Theorem 1. After that, in Sect. \ref{sect:Lagrange} we demostrate some
further applications of the
three--leg forms, establishing the variational and symplectic structures
for Eqs. (\ref{basic eq}) and (\ref{Toda}).

The theorem below provides three-leg forms for all equations of the lists Q
and H (the results for the list A follow from these ones). As it turns out,
in almost all cases it is more convenient to write
the three-leg equation (\ref{3a}) in the multiplicative form
\begin{equation}\label{3m}
\Psi(x,u;\a)/\Psi(x,v;\b)=\Phi(x,y;\a,\b).
\end{equation}
For the list Q the functions $\Psi$ and $\Phi$ corresponding to the
``short'' and to the ``long'' legs, respectively, essentially coincide.
One has in these cases:
\begin{equation}\label{3mF}
\Psi(x,u;\a)=F(X,U;A),\quad \Phi(x,y;\a,\b)=F(X,Y;A-B),
\end{equation}
where some suitable point transformations of the field variables and of the
parameters are introduced: $x=f(X)$, $u=f(U)$, $y=f(Y)$, and $\a=\rho(A)$,
$\b=\rho(B)$.

\begin{theorem}\label{Three-leg forms}
The three--leg forms exist for all equations from Theorem 1.
For the lists Q, H they are listed below.

${\rm (Q1)_{\d=0}}\!\!:$ An additive three--leg form with
$\phi(x,y;\a,\b)=\psi(x,y;\a-\b)$,
\begin{equation}\label{Q10 3leg}
\psi(x,u;\a)=\frac{\a}{x-u}.
\end{equation}

${\rm (Q1)_{\d=1}\!\!:}$ A multiplicative three-leg form with
$\Phi(x,y;\a,\b)=\Psi(x,y;\a-\b)$,
\begin{equation}\label{Q11 3leg}
\Psi(x,u;\a)=\frac{x-u+\a}{x-u-\a}.
\end{equation}

{\rm (Q2):} A multiplicative three-leg form with
$\Phi(x,y;\a,\b)=\Psi(x,y;\a-\b)$,
\begin{equation}\label{Q2 3leg Psi}
\Psi(x,u;\a)=\frac{(X+U+\a)(X-U+\a)}{(X+U-\a)(X-U-\a)},
\end{equation}
where $x=X^2$, $u=U^2$.

${\rm (Q3)_{\d=0}\!\!:}$ A multiplicative three-leg form with
$\Phi(x,y;\a,\b)=\Psi(x,y;\a/\b)$,
\begin{equation}\label{Q30 3leg Psi}
\Psi(x,u;\a)=\frac{\a x-u}{x-\a u}.
\end{equation}
Under the point transformations $x=\exp(2X)$, $\a=\exp(2A)$, etc., there
holds (\ref{3mF}) with
\begin{equation}\label{Q30 3leg F}
\Psi(x,u;\a)=F(X,U;A)=\frac{\sinh(X-U+A)}{\sinh(X-U-A)}.
\end{equation}

${\rm (Q3)_{\d=1}\!\!:}$ A multiplicative three-leg form with (\ref{3mF}),
\begin{equation}\label{Q31 3leg Psi}
\Psi(x,u;\a)=F(X,U;A)=
  \frac{\sinh(X+U+A)\sinh(X-U+A)}{\sinh(X+U-A)\sinh(X-U-A)},
\end{equation}
where $x=\cosh(2X)$, $\a=\exp(2A)$, etc.

{\rm (Q4):} A multiplicative three-leg form with (\ref{3mF}),
\begin{equation}\label{Q4 3leg Psi}
\Psi(x,u;\a)=F(X,U;A)=
  \frac{\sigma(X+U+A)\sigma(X-U+A)}{\sigma(X+U-A)\sigma(X-U-A)},
\end{equation}
where $x=\wp(X)$, $\a=\wp(A)$, etc.

{\rm (H1):} An additive three-leg form
\begin{equation}\label{H1 3leg phi psi}
\psi(x,u;\a)=x+u,\quad \phi(x,y;\a,\b)=\frac{\a-\b}{x-y}.
\end{equation}

{\rm (H2):} A multiplicative three--leg form
\begin{equation}\label{H2 3leg phi psi}
\Psi(x,u;\a)=x+u+\a,\quad \Phi(x,y;\a,\b)=\frac{x-y+\a-\b}{x-y-\a+\b}.
\end{equation}

{\rm (H3):} A multiplicative three--leg form
\begin{eqnarray}
\Psi(x,u;\a) & = & xu+\d\a\;=\;\exp(2X+2U)+\d\a,
   \label{H3 3leg psi}\\
\Phi(x,y;\a,\b) & = & \frac{\b x-\a y}{\a x-\b y}\;=\;
                \frac{\sinh(X-U-A+B)}{\sinh(X-U+A-B)},
   \label{H3 3leg phi}
\end{eqnarray}
where $x=\exp(2X)$, $\a=\exp(2A)$, etc.
\end{theorem}
{\bf Proof.} We start with the list H, for which the situation is
somewhat simpler. Finding the three-leg forms of Eqs. (H1) and
${\rm (H3)}_{\d=0}$ is almost immediate: these equations are
equivalent to
\begin{equation}\label{H1 3leg}
u-v=\frac{\a-\b}{x-y}\qquad {\rm and}\qquad
\frac{u}{v}=\frac{\b x-\a y}{\a x-\b y},
\end{equation}
respectively. For other equations of the list H one uses the following
simple formula which will be also quoted on several occasions later on.
\begin{lemma}\label{lemma hhQQ}
The relation $Q(x,u,v,y;\a,\b)=0$ yields
\begin{equation}\label{H 3leg}
   \frac{h(x,u;\a)}{h(x,v;\b)}=-\frac{Q_v}{Q_u}.
\end{equation}
\end{lemma}

\noindent
{\it Proof.}
\[
   -\frac{h(x,u;\a)}{h(x,v;\b)}= \frac{g(x,u;\a,\b)}{g(x,v;\b,\a)}
    = \frac{QQ_{yv}-Q_yQ_v}{QQ_{yu}-Q_yQ_u}
    = \left.\frac{Q_v}{Q_u}\right|_{Q=0}. \qquad\qed
\]

This lemma immediately yields the three-leg forms for the cases when
$Q_{uv}=0$: then the right--hand side of (\ref{H 3leg})
does not depend on $u,v$ and can therefore be taken as $\Phi(x,y;\a,\b)$,
while $\Psi(x,u;\a):=h(x,u;\a)$. This covers the cases (H2) and (H3). Observe
also that under the condition $Q_{uv}=0$ we have
$G(x,y;\a,\b)=QQ_{uv}-Q_uQ_v=-Q_uQ_v$. Thus, in these cases the function
$\Psi$ associated to the ``short'' legs is nothing but the polynomial $h$,
while the function $\Phi$ associated to the ``long'' legs, is just a ratio
of two linear factors of the quadratic polynomial $G$.

The situation with the list Q is a bit more tricky. The inspection of
formulas (q1), (q2), (q3), (q4) shows that in these cases $h(x,u;\a)$ is a
quadratic polynomial in $u$, and,  similarly, the polynomial $G(x,y;\a,\b)$
is a quadratic polynomial in $y$. We try a linear--fractional ansatz for
$\Psi$, $\Phi$:
%
\begin{equation}\label{PsiPhi factored}
  \Psi(x,u;\a)=\frac{p_+(x,u;\a)}{p_-(x,u;\a)},\quad
  \Phi(x,y;\a,\b)=\frac{s_+(x,y;\a,\b)}{s_-(x,y;\a,\b)},
  \end{equation}
where the functions $p_{\pm}(x,u,\a)$ are linear in $u$, and the functions
$s_{\pm}(x,y;\a,\b)$ are linear in $y$, and
\begin{eqnarray*}
  p_+(x,u;\a)p_-(x,u;\a) & = & h(x,u;\a), \\
  s_+(x,y;\a,\b)s_-(x,y;\a,\b) & = & G(x,y;\a,\b).
\end{eqnarray*}
According to (\ref{dis3}), the coefficients of polynomials
$p_{\pm}(\cdot,u;\cdot)$ and $s_{\pm}(\cdot,y;\cdot,\!\cdot)$
are rational functions of $x$ and $\sqrt{r(x)}$. This justifies
``uniformizing'' changes of variables, namely
\begin{itemize}
\item[] $x=X^2$ in the case (Q2), when $r(x)=x$,
\item[] $x=\cosh(2X)$ in the case (Q3) with $\d=1$, when $r(x)=x^2-1$,
\item[] $x=\wp(X)$ in the case (Q4), when $r(x)=4x^3-g_2x-g_3$.
\end{itemize}
The ansatz (\ref{PsiPhi factored}) turns out to work. For the
cases (Q1)--(Q3) identify the functions $p_{\pm}$ from
(\ref{PsiPhi factored}) with the numerators and denominators of
the expressions listed in (\ref{Q11 3leg})--(\ref{Q31 3leg Psi})
(for the case ${\rm (Q3)}_{\d=0}$ take the fraction (\ref{Q30 3leg Psi}),
i.e. the expression through $x,u$). In the case (Q4) the functions
$p_{\pm}$ linear in $u$ are obtained by dividing the numerators and
denominators of the fractions in (\ref{Q4 3leg Psi}) by
$\sigma^2(X)\sigma^2(U)$. Make similar identifications for the functions
$s_{\pm}$ from (\ref{PsiPhi factored}). So, under the point transformations
used in the formulation of Theorem \ref{Three-leg forms} we have:
\[
p_{\pm}(x,u;\a)=P(X,U;\pm A),\quad s_{\pm}(x,y;\a,\b)=P(X,U;\pm A\mp B),
\]
where
\begin{eqnarray*}
{\rm (Q1)}_{\d=1}: & &  P(x,u;\a)=x-u+\a, \\
{\rm (Q2)}\;\;\;\;\,:  & & P(X,U;\a)=(X+U+\a)(X-U+\a), \\
{\rm (Q3)}_{\d=0}: & &  P(X,U;A)=\sinh(X-U+A), \\
{\rm (Q3)}_{\d=1}: & &  P(X,U;A)=\sinh(X+U+A)\sinh(X-U+A), \\
{\rm (Q4)}\;\;\;\;\,: & &  
   P(X,U;A)=\frac{\sigma(X+U+A)\sigma(X-U+A)}{
                           \sigma^2(X)\sigma^2(U)}.					          
\end{eqnarray*}
A straightforward computation shows that
\begin{eqnarray}
p_+(x,u;\a)p_-(x,v;\b)s_-(x,y;\a,\b)-p_-(x,u;\a)p_+(x,v;\b)s_+(x,y;\a,\b)
\nonumber\\
=\rho(x;\a,\b)Q(x,u,v,y;\a,\b),\qquad\qquad\qquad\qquad
\label{Q vs 3-leg}
\end{eqnarray}
with some factor $\rho$ depending only on $x$. (Obviously,
the left--hand side of the latter equation is linear in $u,v,y$.)
Concretely, we find: $\rho=2$ in the case (${\rm Q1)_{\d=1}}$;
$\rho=4X$ in the case (Q2); $\rho=x$ in the case (${\rm Q3)_{\d=0}}$;
$\rho=2\a\b\sinh(2X)$ in the case (${\rm Q3)_{\d=1}}$; finally,
\[
\rho=\sigma^4(A)\sigma^4(B)\frac{\sigma(B-A)}{\sigma(B+A)}\cdot
\frac{\sigma(2X)}{\sigma^4(X)}
\]
in the case (Q4). So, in all these cases $Q=0$ is equivalent to vanishing
of the left--hand side of (\ref{Q vs 3-leg}), which, in turn, is equivalent
to the multiplicative three--leg formula (\ref{3m}) with the ansatz
(\ref{PsiPhi factored}). Finally, we comment on the origin of the additive
three--leg form of ${\rm (Q1)_{\d=0}}$. Rescaling in (\ref{Q11 3leg}) the
parameters as $\a\mapsto\d\a$, we come to the three--leg form of the
equation (Q1) with a general $\d\neq 0$. Sending $\d\to 0$, we find:
$\Psi=1+2\d\psi+O(\d^2)$, $\Phi=1+2\d\phi+O(\d^2)$, with the functions
$\psi$, $\phi$ from (\ref{Q10 3leg}), and thus we arrive at the additive
formula for the case $\d=0$. \qed

\paragraph{Remark.} The notion of the three--leg equation was formulated
in \cite{BS}. The three--leg forms for Eqs. (Q1), ${\rm
(Q3)}_{\d=0}$, (H1), ${\rm (H3)}_{\d=0}$ were also found there.
The results for (Q2), ${\rm (Q3)}_{\d=1}$, (Q4), (H2), ${\rm
(H3)}_{\d=1}$ are given here for the first time.

\section{Lagrangian structures}\label{sect:Lagrange}

Recall that the three--leg form of Eq. (\ref{basic eq}) imply that the
discrete Toda type equation (\ref{Toda})
holds on the black and on the white sublattices. (Of course, for
multiplicative equations (\ref{3m}) we set $\psi(x,u;\a)=\log\Psi(x,u;\a)$,
$\phi(x,y;\a,\b)=\log\Phi(x,y;\a,\b)$.) We show now that all these Toda
systems may be given a variational (Lagrangian) interpretation. Notice that
only the functions $\phi(x,y;\a,\b)$ related to the ``long'' legs enter
the Toda equations, and that the Toda systems for the cases (H1)--(H3) are
the same as in the cases (Q1), $({\rm Q3})_{\d=0}$.

Consider the point changes of variables $x=f(X)$ listed in Theorem
\ref{Three-leg forms}; in the cases (Q1), (H1), (H2), when no such
substitution is listed, set just $x=X$. For the sake of notational
simplicity, we write in the present section $\psi(x,u;\a)$ for
$\psi(f(X),f(U);\a)$, etc.
\begin{lemma}\label{Observation}
For all equations from Theorem 1 there exist symmetric functions
$L(X,U;\a)=L(U,X;\a)$ and
$\Lambda(X,Y;\a,\b)=\Lambda(Y,X;a,b)$ such that
\begin{eqnarray}
\psi(x,u;\a)=\psi(f(X),f(U);\a) & = & \frac{\partial}{\partial X}L(X,U;\a),\\
\phi(x,y;\a,\b)=\phi(f(X),f(Y);\a,\b) & = &
\frac{\partial}{\partial X}\Lambda(X,Y;\a,\b).
\end{eqnarray}
\end{lemma}
\begin{proof} It is sufficient to notice that the functions
$(\partial/\partial U)\psi(x,u;\a)$ are symmetric with respect
to the permutation $X\leftrightarrow U$, and similarly for $\phi$.
\end{proof}
This observation has the following immediate corollary.
\begin{proposition}
Let $\bigstar$ be the ``black'' subgraph, and let $E(\bigstar)$ be the
set of its (non-oriented) edges. Let the pairs of labels $(\a,\b)$ be
assigned to the edges from $E(\bigstar)$ according to Fig.\,\ref{Fig:star},
so that, e.g., the pair $(\a_1,\a_2)$ corresponds to the edge $(x,x_{12})$.
Then for all equations from Theorem 1 the discrete Toda type
equations (\ref{Toda}) are the Euler--Lagrange
equations for the action functional
\begin{equation}\label{Toda action}
S=\sum_{(X,Y)\in E(\bigstar)} \Lambda(X,Y;\a,\b).
\end{equation}
\end{proposition}
This result could be anticipated, since it is quite natural to
expect from a system consisting of equations on stars to have a
variational origin. Our next result is, on the contrary, somewhat
unexpected, since it gives a sort of a variational interpretation
for the original system consisting of the equations on
quadrilaterals (\ref{basic eq}). This is possible not on arbitrary
quad--graphs but only on special regular lattices. We restrict
ourselves here to the case of the standard square lattice.
For the cases (H1), ${\rm (H3)}_{\d=0}$ (the discrete KdV
equation and the Hirota equation) our results coincide with those
found in \cite{CNP}, for all other equations from Theorem 1 they seem
to be new.

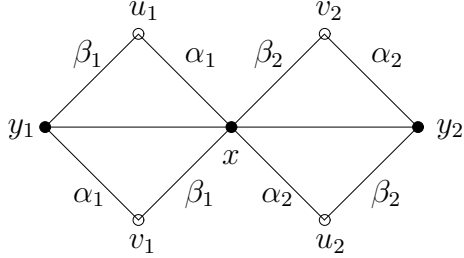
\begin{figure}[htbp]
\begin{center}
\setlength{\unitlength}{0.06em}
\begin{picture}(200,140)(-100,-80)
  \put(0,  0){\circle*{6}} \put(100,0){\circle*{6}}
  \put(-100,  0){\circle*{6}}
  \put( 50,  50){\circle{6}}
  \put( 50,  -50){\circle{6}}
  \put( -50,  50){\circle{6}}
  \put( -50,  -50){\circle{6}}
  \path(-100,0)(100,0)
  \path(-100,0)(-50,50)  \path(-100,0)(-50,-50)
  \path(100,0)(50,50)  \path(100,0)(50,-50)
  \path(0,0)(-50,50)  \path(0,0)(-50,-50)
  \path(0,0)(50,50)  \path(0,0)(50,-50)
  \put(-5,-20){$x$}
  \put(-120,-3){$y_1$}
  \put(110,-3){$y_2$}
  \put(-55,-65){$v_1$}
  \put(-55,60){$u_1$}
  \put(45,-65){$u_2$}
  \put(45,60){$v_2$}
  \put(-25,-40){$\b_1$}  \put(-25,35){$\a_1$}
  \put(-85,-40){$\a_1$}  \put(-85,35){$\b_1$}
  \put(16,-40){$\a_2$}  \put(12,35){$\b_2$}
  \put(75,-40){$\b_2$}  \put(75,35){$\a_2$}
\end{picture}
\caption{Two elementary quadrilaterals of the square lattice}
\label{Fig:lc squares}
\end{center}
\end{figure}

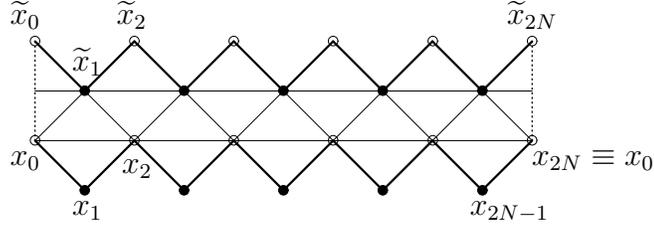
\begin{figure}[htbp]
\begin{center}
\setlength{\unitlength}{0.08em}
\begin{picture}(200,100)(0,-20)
  \put(20,0){\circle*{4}} \put(60,0){\circle*{4}}
  \put(100,0){\circle*{4}} \put(140,0){\circle*{4}}
  \put(180,0){\circle*{4}}
  \put(0,20){\circle{4}} \put(40,20){\circle{4}}
  \put(80,20){\circle{4}} \put(120,20){\circle{4}}
  \put(160,20){\circle{4}} \put(200,20){\circle{4}}
  \put(20,40){\circle*{4}} \put(60,40){\circle*{4}}
  \put(100,40){\circle*{4}} \put(140,40){\circle*{4}}
  \put(180,40){\circle*{4}}
  \put(0,60){\circle{4}} \put(40,60){\circle{4}}
  \put(80,60){\circle{4}} \put(120,60){\circle{4}}
  \put(160,60){\circle{4}} \put(200,60){\circle{4}}
  \path(0,20)(20,40) \path(20,40)(40,20)
  \path(40,20)(60,40) \path(60,40)(80,20)
  \path(80,20)(100,40) \path(100,40)(120,20)
  \path(120,20)(140,40) \path(140,40)(160,20)
  \path(160,20)(180,40) \path(180,40)(200,20)
  \path(0,20)(200,20) \path(0,40)(200,40)
\dottedline{2}(0,20)(0,60) \dottedline{2}(200,20)(200,60)
\thicklines
  \path(0,20)(20,0) \path(20,0)(40,20)
  \path(40,20)(60,0) \path(60,0)(80,20)
  \path(80,20)(100,0) \path(100,0)(120,20)
  \path(120,20)(140,0) \path(140,0)(160,20)
  \path(160,20)(180,0) \path(180,0)(200,20)
  \path(0,60)(20,40) \path(20,40)(40,60)
  \path(40,60)(60,40) \path(60,40)(80,60)
  \path(80,60)(100,40) \path(100,40)(120,60)
  \path(120,60)(140,40) \path(140,40)(160,60)
  \path(160,60)(180,40) \path(180,40)(200,60)
  \put(-10,10){$x_0$} \put(15,-10){$x_1$} \put(35,7){$x_2$}
  \put(175,-10){$x_{2N-1}$}
  \put(200,10){$x_{2N}\equiv x_0$}
  \put(-10,67){$\wx_0$}
  \put(15,48){$\wx_1$}
  \put(33,67){$\wx_2$}
  \put(190,67){$\wx_{2N}$}
\end{picture}
\caption{The Cauchy problem on a zigzag} \label{Fig:Cauchy}
\end{center}
\end{figure}

For the sake of convenience orient the square lattice as in
Fig.\,\ref{Fig:lc squares}.
Denote by $E_1$ ($E_2$) the subset of edges running from south--east
to north--west (resp. from south--west to north--east). Add to the edges of
the original quad--graph the set $E_3$ of the horizontal diagonals of all
elementary quadrilaterals. Let the labels $\alpha$ be assigned to the edges
from $E_1$, and the let the labels $\beta$ be assigned to the edges
from $E_2$. It is natural to assign to the diagonals from $E_3$ the pairs
$(\a,\b)$ from the sides of the correspondent quadrilateral.
\begin{proposition}\label{Lagrangian for 4-point eq}
Solutions of all equations from Theorem 1 are critical for
the following action functional:
\begin{equation}\label{quad action}
{\bf S}=\sum_{(X,U)\in E_1} L(X,U;\a)-\sum_{(X,V)\in E_2} L(X,V;\b)-
\sum_{(X,Y)\in E_3}\Lambda(X,Y;\a,\b).
\end{equation}
\end{proposition}
\begin{proof}
For any vertex $x$, the correspondent Euler--Lagrange equation
relates the vertices of two elementary quadrilaterals, as in
Fig.\,\ref{Fig:lc squares}. Due to Lemma \ref{Observation}, this
equation reads:
\begin{eqnarray*}
\psi(x,u_1;\a_1)-\psi(x,v_1;\b_1)-\phi(x,y_1;\a_1,\b_1)+ & & \\
\qquad\psi(x,u_2;\a_2)-\psi(x,v_2;\b_2)-\phi(x,y_2;\a_2,\b_2) & = & 0.
\end{eqnarray*}
This holds, since (\ref{3a}) is fulfilled on both elementary
quadrilaterals.
\end{proof}
The variational interpretation allows one to find invariant
symplectic structures for reasonably posed Cauchy problems for
Eqs. (\ref{basic eq}). As is well--known, one way to set the
Cauchy problem is to prescribe the values $x$ on the zigzag line
like in Fig.\,\ref{Fig:Cauchy} and to impose periodicity in the
horizontal direction (so that one is dealing with the square
lattice on a cylinder) \cite{CNP}, \cite{FV}. Then Eq. (\ref{basic eq})
defines the evolution in the vertical direction, i.e. the map
$\{x_i\}_{i\in\mathbb Z/2N\mathbb Z}\mapsto
\{\widetilde{x}_i\}_{i\in\mathbb Z/2N\mathbb Z}$.
\begin{proposition}\label{sym struct}
Let the edges $(x_i,x_{i+1})$ carry the labels $\a_i$, and let the
edges $(\wx_i,\wx_{i+1})$ carry the labels $\widetilde{\a}_i$, so
that $\widetilde{\a}_{2k}=\a_{2k-2}$, and
$\widetilde{\a}_{2k-1}=\a_{2k+1}$. Denote
\begin{equation}\label{s}
s(X,U;\a)=
\frac{\partial}{\partial U}\psi(x,u;\a)=
\frac{\partial^2}{\partial X\partial U}L(X,U;\a),
\end{equation}
so that $s(X,U;\a)=s(U,X;\a)$. Then the following relation holds for the
map $\{x_i\}\mapsto \{\widetilde{x}_i\}$:
\begin{equation}\label{sym}
\sum_{i\in\mathbb{Z}/2N\mathbb{Z}}
s(X_{i},X_{i+1};\a_i)dX_{i}\wedge dX_{i+1}=
\sum_{i\in\mathbb{Z}/2N\mathbb{Z}}
s(\wX_{i},\wX_{i+1};\widetilde{\a}_i)d\wX_{i}\wedge d\wX_{i+1}\;.
\end{equation}
\end{proposition}
\begin{proof} We use the argument methodologically close to \cite{MPS}.
Consider the action functional $\mathbb S$ defined by the same formula as
(\ref{quad action}) but with the summations restricted to the edges depicted
in Fig.\,\ref{Fig:Cauchy}, and restricted to such fields $X$ which satisfy
the equation (\ref{basic eq}) on each elementary quadrilateral. Consider
the differential
\[
d\mathbb S=\sum_{{\rm all\;vertices}\;x}\frac{\partial\mathbb S}{\partial X}dX.
\]
Due to Proposition \ref{Lagrangian for 4-point eq}, the expression
$\partial\mathbb S/\partial X$ vanishes for all vertices where six edges
meet, i.e. for all vertices except $x_{2k+1}$ and $\wx_{2k}$. So, we find:
\begin{eqnarray*}
d\mathbb S & = & \sum_{k\in\mathbb{Z}/N\mathbb{Z}} \frac{\partial
L(X_{2k},X_{2k+1};\a_{2k})}{\partial X_{2k+1}}dX_{2k+1}-
\frac{\partial L(X_{2k+1},X_{2k+2};\a_{2k+1})}{\partial
X_{2k+1}}dX_{2k+1}\\ &  & -\frac{\partial
L(\wX_{2k-1},\wX_{2k};\widetilde{\a}_{2k-1})} {\partial
\wX_{2k}}d\wX_{2k} +\frac{\partial
L(\wX_{2k},\wX_{2k+1};\widetilde{\a}_{2k})} {\partial
\wX_{2k}}d\wX_{2k}.
\end{eqnarray*}
Differentiating this 1--form and taking into account $d^2\mathbb S=0$
yields (\ref{sym}).
\end{proof}

\paragraph{Remarks.} 1) Notice that only the functions $\psi$ related to the
``short'' legs are present in formula (\ref{sym}).

2) One is tempted to interpret formula (\ref{sym}) as symplecticity of
 the map $\{X_i\}\mapsto\{\wX_i\}$. However, the 2--form (\ref{sym}) is
 degenerate. One finds a genuine symplectic form, if one considers a
 quasi-periodic initial value problem instead of a periodic one, and extends
 the phase space by the correspondent monodromies (cf. \cite{FV}, \cite{K}
 for the case ${\rm (H3)}_{\d=0}$ (the Hirota equation)). Actually, the
 formula (\ref{sym}) means the invariance of a degenerate
 2--form only when $\widetilde{\a}_i=\a_i$. This is easy to achieve
 by setting all $\a_i=\a$, or, more generally, all $\a_{2k}=\a$
 and all $\a_{2k+1}=\b$.
 \vspace{5pt}

We close this section with a local form of Proposition  \ref{sym struct},
also based on the three--leg form of Eqs. (\ref{basic eq}).
The particular case of the Hirota equation was given in \cite{K}.
\begin{proposition}\label{elementary omega}
Associate the two--form $\omega(e)=s(X,U;\a)dX\wedge dU$
to every (oriented) edge $e=(x,u)$ of the quad--graph.
Then Eq. (\ref{basic eq}) with the three--leg form (\ref{3a})
implies that the sum of $\omega$'s along the boundary of any
elementary quadrilateral, and therefore along any cycle homotopic
to zero, vanishes.
\end{proposition}
\begin{proof}
Differentiating (\ref{3a}) and wedging the result with $dX$, we
have:
\[
\frac{\partial}{\partial U}\psi(x,u;\a)dX\wedge dU+
\frac{\partial}{\partial V}\psi(x,v;\b)dV\wedge dX=
\frac{\partial}{\partial Y}\phi(x,y;\a)dX\wedge dY.
\]
On the other hand, starting with the three--leg equation centered
at $y$ (or, in other words, flipping $x\leftrightarrow y$,
$u\leftrightarrow v$), we arrive at
\[
\frac{\partial}{\partial V}\psi(y,v;\a)dY\wedge dV+
\frac{\partial}{\partial U}\psi(y,u;\b)dU\wedge dY=
\frac{\partial}{\partial X}\phi(y,x;\a)dY\wedge dX.
\]
Adding these two equations, and taking into account Lemma
\ref{Observation} and the notation (\ref{s}), we come to the
statement of Proposition.
\end{proof}

The statement of Proposition \ref{sym struct}
is a particular case of Proposition \ref{elementary
omega}, since the boundary of the domain in Fig.\,\ref{Fig:Cauchy}
(a cylindrical strip) is homotopic to zero.


\section{Relation to B\"acklund transformations}\label{sect:Backlund}

In this section we interpret the integrable quad-graph equations of
Theorem 1 as nonlinear superposition principles (NSP) of B\"acklund
transformations for the KdV--type equations.

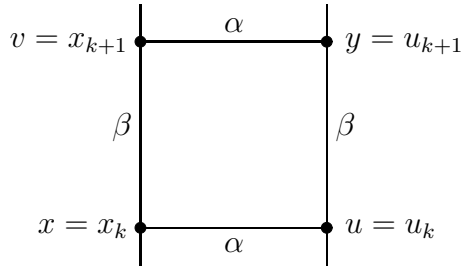
\begin{figure}[htbp]
\begin{center}
\setlength{\unitlength}{0.06em}
\begin{picture}(200,140)(-50,-20)
  \put(100,  0){\circle*{6}} \put(0  ,100){\circle*{6}}
  \put(  0,  0){\circle*{6}}  \put(100,100){\circle*{6}}
  \put( 0,  0){\line(1,0){100}}
  \put( 0,100){\line(1,0){100}}
  \put(  0, -20){\line(0,1){140}}
  \put(100, -20){\line(0,1){140}}
  \put(-55,-3){$x=x_k$}
  \put(110,-3){$u=u_k$}
  \put(110,97){$y=u_{k+1}$}
  \put(-70,97){$v=x_{k+1}$}
  \put(45,-13){$\a$}
  \put(45,105){$\a$}
  \put(-15,50){$\b$}
  \put(105,50){$\b$}
\end{picture}
\caption{To the construction of B\"acklund transformations}\label{Fig:Back}
\end{center}
\end{figure}

First of all, we show how B\"acklund transformations themselves
can be derived from our equations. Towards this aim, consider Eq.
(\ref{basic eq}) on one vertical strip of the standard square
lattice:
\[
S=\{0,1\}\times{\mathbb Z}.
\]
The fields on $S_0=\{0\}\times{\mathbb Z}$ will be denoted by $x_k$, while the
fields on $S_1=\{1\}\times{\mathbb Z}$ will be denoted by $u_k$. So, a single
square of the strip $S$ looks as on Fig.\,\ref{Fig:Back}.
Suppose now that $x_k=x(k\epsilon)$, $k\in\mathbb Z$, where
$x(\xi)$ is a smooth function, and similarly for $u_k$. In
particular, this means that one has to set in Eq. (\ref{basic eq})
$v=x+\epsilon x_{\xi}+O(\epsilon^2)$ and $y=u+\epsilon
u_{\xi}+O(\epsilon^2)$. If, in addition, the parameter $\b$ is
chosen properly, then Eq. (\ref{basic eq}) approximates in the
limit $\epsilon\to 0$ some differential equation which relates the
functions $x(\xi)$ and $u(\xi)$. For the equations of the list Q
the result is the most straightforward.
\begin{proposition}
Set in Eq. (Q1) $\b=\epsilon^2/2$, in Eq.
(Q2) $\b=\epsilon^2/4$, in Eq. (Q3) $\b=1-\epsilon^2/2$, and in
Eq. (Q4) $\b=\wp(\epsilon^2)$. Then in the limit $\epsilon\to 0$ just 
described these equations tend to
\begin{equation}\label{KN Back}
x_{\xi}u_{\xi}=h(x,u;\a),
\end{equation}
with the correspondent polynomials $h$ listed in the
formulas (q1), (q2), (q3), (q4) of Proposition \ref{st:rh}. 
\end{proposition}
{\bf Proof.} The statement is almost obvious in the
cases (Q1), (Q2). To see that it holds for Eq. (Q3), the latter 
can be first rewritten as
\begin{eqnarray*}
 \lefteqn{\b(\a^2-1)(x-v)(u-y)=}\\
 && \a(\b^2-1)(xv+uy)+(1-\b)(\a^2+\b)(xy+uv)\\
 &&+(\d^2/4\a\b)(\a^2-\b^2)(\a^2-1)(\b^2-1).
\end{eqnarray*}
If $\b=1-\epsilon^2/2$, then the above equation approximates
\[
 (\a^2-1)x_{\xi}u_{\xi}=
 -\a(x^2+u^2)+(\a^2+1)xu-(\d^2/4\a)(\a^2-1)^2.
\]
Finally, in the most intricate case (Q4) one starts by rewriting the 
equation as
\begin{eqnarray*}
\lefteqn{
\frac{\bar{a}_2-a_2}{2a_0}\, (x-v)(u-y)=
xuvy+\frac{a_1}{a_0}\,(xuv+uvy+vyx+yxu)}\\
&&+\frac{\bar{a}_2+a_2}{2a_0}\,(x+v)(u+y)+\frac{\tilde{a}_2}{a_0}\,(xv+uy)
+\frac{a_3}{a_0}\,(x+u+v+y)+\frac{a_4}{a_0}\,.
\end{eqnarray*}
From expressions for the coefficients $a_i$ given in Theorem 
\ref{th:list}, it is easy to see that if $\b=\wp(\epsilon^2)\sim\epsilon^{-4}$,
so that $b=\wp'(\epsilon^2)\sim -2\epsilon^{-6}$, then the left--hand
side of the above equation tends to $ax_{\xi}u_{\xi}$, while the right--hand
side tends to
\begin{eqnarray*}
\lefteqn{x^2u^2-2\a xu(x+u)-\Big(2\a^2-\frac{g_2}{2}\Big)xu+\a^2(x^2+u^2)}\\
&&+\Big(g_3+
\frac{g_2}{2}\a\Big)(x+u)+\Big(\frac{g_2^2}{16}+g_3\a\Big)=ah(x,u;\a).
\end{eqnarray*}
This proves the proposition. \qed

Eq. (\ref{KN Back}), read as a Riccati equation for $u$ with the
coefficients dependent on $x$, describes a transformation
$x\mapsto u$, which turns out \cite{A1} to be a {\itbf B\"acklund
transformation} for the Krichever--Novikov equation \cite{KN}:
\begin{equation}\label{KN}
x_t=x_{\xi\xi\xi}-\frac{3}{2x_{\xi}}(x_{\xi\xi}^2-r(x)),
\end{equation}
with the polynomial $r(x)$ being the discriminant of $h(x,u;\a)$.
In other words, if $x$ is a solution of (\ref{KN}), and $u$ is
related to $x$ by (\ref{KN Back}), then $u$ is also a solution of
(\ref{KN}). It should be noticed that, in turn, the partial
differential equation (\ref{KN}) may be derived from (\ref{KN
Back}), either through a sort of continuous limit, or as a higher
symmetry. In any way, it would be fair to say that the whole
theory of the equation (\ref{KN}) and its B\"acklund
transformations (\ref{KN Back}) is contained in the correspondent
quad-graph equation (\ref{basic eq}) from the list Q. To complete
the picture, we demonstrate that all equations (\ref{basic eq})
listed in Theorem 1, in turn, can be interpreted as nonlinear
superposition principles for the B\"acklund transformations. To
this end consider the system of four differential equations of the
type (\ref{KN Back}) corresponding to four sides of the
quadrilateral  on Fig.\,\ref{Fig:quadrilateral}:
\begin{align}
   x_{\xi}u_{\xi}=h(x,u;\a), & \qquad u_{\xi}y_{\xi}=h(u,y;\b),
   \label{4 Back 1}\\
   x_{\xi}v_{\xi}=h(x,v;\b), & \qquad v_{\xi}y_{\xi}=h(v,y;\a).
   \label{4 Back 2}
\end{align}
We will consider it for functions $h$ corresponding to all cases
listed in Theorem 1, except for those two when $h$ actually does
not depend on parameters, namely (H1) with $h(x,u)=1$, and ${\rm
(H3)}_{\d=0}$ with $h(x,u)=xu$. In all other cases the system
(\ref{4 Back 1}), (\ref{4 Back 2}) makes perfect sense and its
consistency is quite nontrivial.

\begin{proposition}\label{cor:chain}
The equation $Q(x,u,v,y;\a,\b)=0$ is a sufficient condition for
the consistency of the system of differential equations (\ref{4
Back 1}), (\ref{4 Back 2}). Moreover, it is compatible with this
system, i.e.
\begin{equation}
Q_{\xi}|_{Q=0}=0.
\end{equation}
\end{proposition}
\begin{proof} The consistency condition of the differential equations
(\ref{4 Back 1}),(\ref{4 Back 2}) reads:
\begin{equation}\label{hh-hh}
\cQ:=h(x,u;\a)h(v,y;\a)-h(x,v;\b)h(u,y;\b)=0.
\end{equation}
The left--hand side $\cQ$ of this equation is a polynomial of
degree 1 in each variable for the equations of the list H, and of
degree 2 in each variable for the equations of the list Q. In the
former case it is directly seen that Eq. (\ref{hh-hh}) exactly
coincides with $Q(x,u,v,y;\a,\b)=0$. In the latter case it can be
shown that the polynomial $Q$ divides the left--hand side of
(\ref{hh-hh}). More precisely, it is verified that, up to a
constant factor, $\cQ=QP$, where in the cases (Q1), (Q2) one has
$P=Q|_{\b\to -\b}$, in the case (Q3) one has $P=Q|_{\b\to 1/\b}$,
and in the case (Q4) one has $P=Q|_{(\b,b)\to (\b,-b)}$. This
proves the first statement of proposition. The second one reads as
\[
   (Q_xx_{\xi}+Q_uu_{\xi}+Q_vv_{\xi}+Q_yy_{\xi})|_{Q=0}=0.
\]
We will prove that actually both terms $Q_xx_{\xi}+Q_yy_{\xi}$ and
$Q_uu_{\xi}+Q_vv_{\xi}$ vanish separately. For example,
\[
   (Q_uu_{\xi}+Q_vv_{\xi})|_{Q=0}
  =\frac{1}{x_{\xi}}\left.\big(Q_uh(x,u;\a)+Q_vh(x,v;\b)\big)\right|_{Q=0}=0.
\]
The last step follows from Lemma \ref{lemma hhQQ}.
\end{proof}

In the two cases when the system (\ref{4 Back 1}), (\ref{4 Back
2}) becomes trivial, there still exist B\"acklund transformations
of a different kind, such that the equation $Q=0$ serves as their
NSP. Namely, in the cases (H1), ${\rm (H3)}_{\d=0}$ the following
equations come to replace (\ref{KN Back}):
\begin{eqnarray}
  x_{\xi}+u_{\xi} & = & (x-u)^2+\a,  \label{KdV Back}\\ \nonumber\\
  \frac{x_{\xi}}{x}+\frac{u_{\xi}}{u} & = & \frac{\a}{2}\left(\frac{x}{u}+
  \frac{u}{x}\right),\label{mKdV Back}
\end{eqnarray}
respectively. The second of these equations is probably better known in the
coordinates $x=\exp(X)$, $u=\exp(U)$:
\begin{equation}\label{MKdV Back}
  X_{\xi}+U_{\xi}=\a\cosh(X-U).
\end{equation}
Eq. (\ref{KdV Back}) defines a B\"acklund transformation for the
potential KdV equation
\begin{equation}\label{pot KdV}
  x_t=x_{\xi\xi\xi}-6x_{\xi}^2.
\end{equation}
Similarly, Eq. (\ref{MKdV Back}) defines a B\"acklund
transformation for the potential MKdV equation
\begin{equation}\label{pot MKdV}
  X_t=X_{\xi\xi\xi}-2X_{\xi}^3,
\end{equation}
or, alternatively, for the sinh--Gordon equation
which belongs to the same hierarchy.

\section{Conclusions and perspectives.}\label{sect:conclusions}

{\bf Three--dimensional consistency as integrability criterium; Yang--Baxter
maps.} One of the traditional but rather {\it ad hoc} definitions of the 
integrability of two--dimensional systems is based on the notion of
the zero--curvature representation. For a system on a quad--graph
consisting of equations (\ref{basic eq}) with fields associated to
the {\it vertices} of the elemntary quadrilateral on Fig. 
\ref{Fig:quadrilateral}, the zero curvature representation is usually
encoded in the formula like
\[
L(y,u,\b;\lambda)L(u,x,\a;\lambda)=L(y,v,\a;\lambda)L(v,x,\b;\lambda),
\]
where $\lambda$ is the spectral parameter, so that the matrices $L$
take values in some loop group. In \cite{BS} and independently in \cite{N}
it was demonstrated how to derive the zero curvature representation from
the three--dimensional consistency. It should be mentioned, however,
that to assign fields to the vertices is not the only possibility.
Another large class of two--dimensional systems on quad--graphs build
those with the fields assigned to the {\it edges}, see Fig. \ref
{Fig:quadrilateral edges}. 
\begin{figure}[htbp]
\begin{minipage}[t]{150pt}
\setlength{\unitlength}{0.06em}
\begin{picture}(200,140)(-50,-20)
  \put( 0,  0){\line(1,0){100}}
  \put( 0,100){\line(1,0){100}}
  \put(  0, 0){\line(0,1){100}}
  \put(100, 0){\line(0,1){100}}
  \put(47,-13){$\a$} \put(47,8){$a$}
  \put(47,105){$\a$} \put(47,84){$a_2$}
  \put(-13,47){$\b$} \put(8,47){$b_1$}
  \put(105,47){$\b$} \put(84,47){$b$}
\end{picture}
\caption{An elementary quadrilateral; both fields and labels are assigned 
to edges}
\label{Fig:quadrilateral edges}
\end{minipage}\hfill
\begin{minipage}[t]{200pt}
\setlength{\unitlength}{0.08em}
\begin{picture}(200,170)(-50,-20)
  \put( 0,  0){\line(1,0){100}}
  \put( 0,100){\line(1,0){100}}
  \put(50,130){\line(1,0){100}}
  \multiput(50,30)(20,0){5}{\line(1,0){15}}
  \put(  0, 0){\line(0,1){100}}
  \put(100, 0){\line(0,1){100}}
  \put(150,30){\line(0,1){100}}
  \multiput(50,30)(0,20){5}{\line(0,1){15}}
  \put(  0,100){\line(5,3){50}}
  \put(100,100){\line(5,3){50}}
  \put(100,  0){\line(5,3){50}}
  \multiput(50,30)(-16.67,-10){3}{\line(-5,-3){12}}
     \put(40,-11){$a$}
     \put(78,20){$a_2$}
     \put(60,105){$a_3$}
     \put(90,135){$a_{23}$}
     \put(130,6){$b$}
     \put(20,25){$b_1$}
     \put(113,117){$b_3$}
     \put(14,120){$b_{13}$}
     \put(155,75){$c$}
     \put(35,75){$c_1$}
     \put(105,50){$c_2$}
     \put(-17,50){$c_{12}$}
\end{picture}
\caption{Three--dimensional consistency; fields assigned to edges}
\label{Fig:cube edges}
\end{minipage}\hfill
\end{figure}
In this situation it is natural to assume that each elementary quadrilateral
carries a map $R:\cX^2\mapsto\cX^2$, where $\cX$ is the space where the
fields $a,b$ take values, so that $(a_2,b_1)=R(a,b;\a,\b)$. The question
on the three--dimensional consistency of such maps is also legitimate and,
moreover, began to be studied recently. The corresponding property can be
encoded in the formula
\begin{equation}\label{YB map}
R_{23}\circ R_{13}\circ R_{12}=R_{12}\circ R_{13}\circ R_{23},
\end{equation}
where each $R_{ij}:\cX^3\mapsto\cX^3$ acts as the map $R$ on the factors 
$i,j$ of the cartesian product $\cX^3$ and acts identically on the
third factor. This equation should be understood as follows. The fields $a,b$
are supposed to be attached to the edges parallel to the 1st and the 2nd
coordinate axes, respectively. Additionally, consider the fields $c$ attached 
to the edges parallel to the 3rd coordinate axis. Then the left--hand side
of (\ref{YB map}) corresponds to the chain of maps along the three hind
faces of the cube on Fig. \ref{Fig:cube edges}:
\[
(a,b)\mapsto(a_2,b_1),\quad (a_2,c)\mapsto(a_{23},c_1),\quad (b_1,c_1)\mapsto
(b_{13},c_{12}),
\]
while its right--hand side corresponds to the chain of the maps along the
three front faces of the cube:
\[
(b,c)\mapsto(b_3,c_2),\quad (a,c_2)\mapsto(a_3,c_{12}),\quad (a_3,b_3)\mapsto
(a_{23},b_{13})
\]
So, Eq. (\ref{YB map}) assures that two ways of obtaining
$(a_{23},b_{13},c_{12})$ from the initial data $(a,b,c)$ lead to the same
results. The maps with this property were introduced by Drinfeld \cite{D}
under the name of ``set--theoretical solutions of the Yang--Baxter equation'', 
an alternative name is ``Yang--Baxter maps'' used by Veselov in the
recent study \cite{V2}, see also references therein. Under some circumstances,
systems with the fields on vertices can be regarded as systems with
the fields on edges or vice versa (this is the case, e.g., for the systems 
(Q1), ${\rm (Q3)_{\d=0}}$, (H1), ${\rm (H3)_{\d=0}}$ of our list, for which 
the variables $X$ enter only in combinations like $X-U$ for edges $(x,u)$), 
but in general the two classes of systems should be considered as 
different. The notion of the zero curvature
representation makes perfect sense for Yang--Baxter maps: such a map can 
be called integrable, if it is equivalent to
\[
L(b,\b;\lambda)L(a,\a;\lambda)=L(c,\a;\lambda)L(d,\b;\lambda).
\]
The problem of integrability of Yang--Baxter maps in the sense of existence 
of a zero--curvature representation is under current investigation \cite{SV}. 
Also the problem of classification of Yang--Baxter maps, like the one achieved 
in the present paper, is of a great importance and interest.
 \medskip

A different direction for the development of the ideas of the
present paper constitute {\it quantum systems}, or, more generally,
{\it systems with non--commutative variables}. To remain in the frame
of the present paper, these are systems (\ref{basic eq}), where the fields
$(x,u,v,y)$ take values in an arbitrary associative (not necessary commutative)
algebra with a unit. It turns out that
the notion of the three--dimensional consistency can be
formulated also for such non--commutative systems. Also the derivation
of the zero curvature representation can be extended to the 
non--commutative framework \cite{BS2}.
 \medskip

It should be mentioned that in the area of the three--dimensional
consistency of classical systems there also remains a number of
interesting open problems. For instance, one of the assumptions
under which the classification was carried out in the present
paper, was less natural, namely the tetrahedron condition. As we
pointed out in Sect. \ref{sect:solv1}, there exist
three--dimensionally consistent equations without the tetrahedron
property, however all examples we are aware of are trivial (linear
or linearizable):
\[
Q(x,u,v,y)=x+y-u-v=0\quad {\rm or} \quad Q(x,u,v,y)=xy-uv=0,
\]
or those obtained from these two by the action of a M\"obius transformation
on all variables. These examples do not contain parameters, and thus the
three--dimensional consistency does not give a zero curvature representation
with a spectral parameter for them (their integrability is anyway obvious).
It would be interesting to find out whether there exist nontrivial examples
violating the tetrahedron property.

There is also a vast field of {\it multi--field} integrable
equations on quad--graphs. Existing examples indicate that their
study is very promising.
 \medskip
 
{\bf Four--dimensional consistency of three--dimensional systems.} 
Ve\-ry promising is also the application of the consistency approach
to the three--dimensional integrability. The role of an {\it ad hoc}
definition of integrability, played in two dimensions by the zero curvature
representation, now goes to the so called {\it local Yang--Baxter equation},
introduced by Maillet and Nijhoff \cite{MN2}. The role of the transition
matrices from the zero curvature representation is played in this novel
structure by certain tensors attached to the elementary two--dimensional
plaquettes of the three--dimensional lattice. There exist a number of results
on finding  this sort of structure for some three--dimensional integrable
systems \cite{Ko, Ka, KKS}. It would be desirable to relate this {\it ad hoc}
notion of integrability to some constructive one. In the spirit of the present
paper, this constructive notion should be the {\it four--dimensional
consistency}. 

In the three--dimensional context there are {\it a priori} many kinds of 
systems, according to where the fields are defined: on the vertices, on the 
edges, or on the elementary squares of the cubic lattice. Consider first the 
situation when the fields are sitting on the the elementary squares. Attach 
the fields $a$, $b$, $c$ to the two--dimensional faces parallel to the
coordinate planes $12$, $23$, $13$, 
respectively, so that $a$, $b$, $c$ are sitting on the bottom, the
left and the front faces of a cube Fig. \ref{fig.cube}, and $a_3$, $b_1$, 
$c_2$ on the top, the right and the back faces. The system under consideration
is a map $S:\cX^3\mapsto\cX^3$ attached to the cube, so that 
$S(a,b,c)=(a_3,b_1,c_2)$. The condition of the four--dimensional consistency
of such a map can be encoded in the formula
\begin{equation}\label{tetrahedron eq}
S_{134}\circ S_{234}\circ S_{124}\circ S_{123}=
S_{123}\circ S_{124}\circ S_{234}\circ S_{134}.
\end{equation}
This equation should be understood as follows. Additionally to the fields
$a$, $b$, $c$, consider the fields $d$, $e$, $f$, attached to the
two--dimensional faces parallel to the coordinate planes $24$, $14$, $34$ 
of the four--dimensional hypercubic lattice. 
Each map of the type $S_{ijk}$ in 
(\ref{tetrahedron eq}) is a map on $\cX^6(a,b,c,d,e,f)$ acting as $S$ on 
the factors of the cartesian product $\cX^6$ corresponding to the variables
sitting on the faces parallel to the planes $ij$, $jk$, $ik$, and acting 
trivially on the other three factors. Thus the left--hand side of 
(\ref{tetrahedron eq}) corresponds to the chain of maps
\begin{eqnarray*}
(a,b,c)\mapsto(a_3,b_1,c_2), && (a_3,d,e)\mapsto(a_{34},d_1,e_2),\\
(b_1,d_1,f)\mapsto(b_{14},d_{13},f_2), &&
(c_2,e_2,f_2)\mapsto(c_{24},e_{23},f_{21}),
\end{eqnarray*}
while the right--hand side of (\ref{tetrahedron eq}) corresponds to the chain 
of maps 
\begin{eqnarray*}
(c,e,f)\mapsto(c_4,e_3,f_1), && (b,d,f_1)\mapsto(b_4,d_3,f_{12}),\\
(a,d_3,e_3)\mapsto(a_4,d_{13},e_{23}), &&
(a_4,b_4,c_4)\mapsto(a_{34},b_{14},c_{24}).
\end{eqnarray*}
Eq. (\ref{tetrahedron eq}) expresses then the fact that two different ways
of obtaining the data $(a_{34},b_{14},c_{24},d_{13},e_{23},f_{12})$ from the 
initial data $(a,b,c,d,e,f)$ lead to identical results. This equation is known 
in the literature as the {\it functional tetrahedron equation} \cite{Ko, KKS}.
(Note that the standard notation used in the literature on the tetrahedron
equation is different: the indices $1\le \alpha,\beta,\gamma\le 6$ of
$S_{\alpha\beta\gamma}$ numerate the two--dimensional coordinate planes.) 
The paper \cite{KKS} contains also a list of solutions of this equation with
$\cX=\mathbb C$, possessing local Yang--Baxter representations with a certain 
ansatz for the participating tensors. One of the most remarkable examples
is the {\it star--triangle map}:
\begin{equation}\label{ST map}
a_3=-\frac{a}{ab-bc-ca}\,,\quad b_1=-\frac{b}{ab-bc-ca}\,,\quad 
c_2=-\frac{c}{ab-bc-ca}\,.
\end{equation}
(Usually this equation is written in a more symmetric form which is obtained
by changing $c\mapsto -c$, with all plus signs in all denominators; however 
in that form the map does not satisfy Eq. (\ref{tetrahedron eq}).)
See also \cite{Ka} for an alternative local Yang--Baxter representation
for this system. In \cite{S1, S2} an important solution of the functional 
tetrahedron equation with $\cX={\mathbb C}^2$ was introduced and studied 
in detail. There seems to be no method available for {\it deriving} the
local Yang--Baxter representation for a given map satisfying the functional
tetrahedron equation.
 \medskip

\begin{figure}[htbp]
\begin{center}
\setlength{\unitlength}{0.07em}
\begin{picture}(200,220)(-100,-90)
 \drawline(\xyx,\xyy)(\yx,\yy)(\yzx,\yzy)
   (\zx,\zy)(\xzx,\xzy)(\xx,\xy)(\xyx,\xyy)(\xyzx,\xyzy)(\yzx,\yzy)
  \drawline(\xyzx,\xyzy)(\xzx,\xzy)
  \dashline{4}(\xx,\xy)(0,0)(\zx,\zy)\dashline{4}(0,0)(\yx,\yy)
 \drawline(\XYX,\XYY)(\YX,\YY)(\YZX,\YZY)
   (\ZX,\ZY)(\XZX,\XZY)(\XX,\XY)(\XYX,\XYY)(\XYZX,\XYZY)(\YZX,\YZY)
  \drawline(\XYZX,\XYZY)(\XZX,\XZY)
  \dashline{4}(\XX,\XY)(-\ox,-\oy)(\ZX,\ZY)\dashline{4}(-\ox,-\oy)(\YX,\YY)
  \dashline{2}(0,0)(-\ox,-\oy)
  \dashline{2}(\xx,\xy)(\XX,\XY)
  \dashline{2}(\yx,\yy)(\YX,\YY)
  \dashline{2}(\zx,\zy)(\ZX,\ZY)
  \dashline{2}(\xyx,\xyy)(\XYX,\XYY)
  \dashline{2}(\xzx,\xzy)(\XZX,\XZY)
  \dashline{2}(\yzx,\yzy)(\YZX,\YZY)
  \dashline{2}(\xyzx,\xyzy)(\XYZX,\XYZY)
  \put(-45,-15){$x$}
  \put(0,-15){$x_1$}
  \put(-14,5){$x_2$}
  \put(-50,20){$x_3$}
  \put(55,3){$x_{12}$}
  \put(-5,20){$x_{13}$}
  \put(-25,50){$x_{23}$}
  \put(55,43){$x_{123}$}
  \put(-137,-85){$x_4$}
  \put(5,-83){$x_{14}$}
  \put(-12,-40){$x_{24}$}
  \put(-141,50){$x_{34}$}
  \put(139,-28){$x_{124}$}
  \put(6,66){$x_{134}$}
  \put(-10,108){$x_{234}$}
  \put(139,108){$x_{1234}$}
\end{picture}
\caption{Hypercube}
\end{center}
\label{hypercube}
\end{figure}
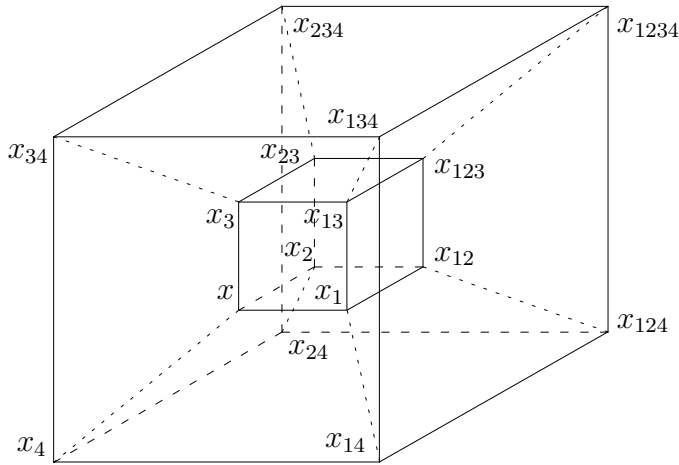

Further, consider three--dimensional systems with the fields sitting
on the vertices. In this case each elementary cube carries just one 
equation
\begin{equation}\label{3D eq}
Q(x,x_1,x_2,x_3,x_{12},x_{23},x_{13},x_{123})=0, 
\end{equation}
relating the fields in all its vertices. Such an equation should be solvable
for any of its arguments in terms of the other seven ones. The 
four--dimensional consistency of such equations is defined as follows.
\begin{itemize}
\item Starting with the initial data $x$, $x_i$ $(1\le i\le
4)$, $x_{ij}$ $(1\le i<j\le 4)$, the equation (\ref{3D eq}) allows us to
uniquely determine all fields $x_{ijk}$ $(1\le i<j<k\le 4)$. Then
we have {\it four} different possibilities to find $x_{1234}$,
corresponding to four three--dimensional cubic faces adjacent to
the vertex $x_{1234}$ of the four--dimensional hypercube, see
Fig.\,\ref{hypercube}. All four values actually coincide.
\end{itemize}
\xx=-35 \xy=-20 \yx=50 \yy=0 \zx=0 \zy=50 \MM=15 \MD=5
 \Sum
 \calcshift(\ox,\xyzx) \calcshift(\oy,\xyzy)
 \calc(\XX,\xx,\ox)      \calc(\XY,\xy,\oy)
 \calc(\YX,\yx,\ox)      \calc(\YY,\yy,\oy)
 \calc(\ZX,\zx,\ox)      \calc(\ZY,\zy,\oy)
 \calc(\XYX,\xyx,\ox)    \calc(\XYY,\xyy,\oy)
 \calc(\XZX,\xzx,\ox)    \calc(\XZY,\xzy,\oy)
 \calc(\YZX,\yzx,\ox)    \calc(\YZY,\yzy,\oy)
 \calc(\XYZX,\xyzx,\ox)  \calc(\XYZY,\xyzy,\oy)
So, one can consistently impose equations (\ref{3D eq}) on all
three--dimen\-sional cubes of the lattice ${\mathbb Z}^4$. 
It is tempting to accept the four--dimensional consistency of equations of
the type (\ref{3D eq}) as the constructive definition of their integrability. 
It will be very important to solve the correspondent classification problem.
We expect that this definition will allow one to derive the local 
Yang--Baxter representation, as a replacement for the zero curvature 
representation characteristic for the two--dimensional integrability. 

We give here some examples. Consider the equation
\begin{equation}\label{8}
\frac{(x_1-x_3)(x_2-x_{123})}{(x_3-x_2)(x_{123}-x_1)}=
\frac{(x-x_{13})(x_{12}-x_{23})}{(x_{13}-x_{12})(x_{23}-x)}.
\end{equation}
This equation appeared for the first time in \cite{NS}, \cite{KS}, along 
with a geometric interpretation. It is not difficult to see that Eq. (\ref{8})
admits a symmetry group $D_8$ of the cube. This equation can be uniquely
solved for a field at an arbitrary vertex of a three--dimensional cube,
provided the fields at other seven vertices are known. 
The fundamental fact not mentioned in \cite{KS} is:
\begin{itemize}
\item Eq. (\ref{8}) is four--dimensionally consistent in the above sense. 
\end{itemize}
This is closely related (in fact, almost synonymous) to the functional 
tetrahedron equation for the star--triangle map (\ref{ST map}), since the 
plaquette variables $(a,b,c)$ of the latter can be ``factorized'' into 
combinations of vertex variables $x$ of Eq. (\ref{8}). More precisely, given 
a solution of (\ref{8}) and setting
\[
a=\frac{x_{12}-x}{x_1-x_2}\,,\quad b=\frac{x_{23}-x}{x_2-x_3}\,,\quad
c=\frac{x_{13}-x}{x_1-x_3}\,
\]
we arrive at a solution of (\ref{ST map}).

A different ``factorization'' of the plaquette variables into the vertex ones
leads to another remarkable three--dimensional system known as the {\it 
discrete BKP equation} \cite{M}, \cite{KS}. For any solution
$x:{\mathbb Z}^4\mapsto\mathbb C$ of (\ref{8}), define a function
$\tau:{\mathbb Z}^4\mapsto\mathbb C$ by the equations
\begin{equation}
\frac{\tau_i\tau_j}{\tau\tau_{ij}}=\frac{x_{ij}-x}{x_i-x_j},\quad i<j.
\end{equation}
Eq. (\ref{8}) assures that this can be done in an essentially unique way
(up to initial data on coordinate axes whose influence is a trivial scaling
of the solution). The function $\tau$ satisfies on any three--dimensional
cube the discrete BKP equation:
\begin{equation}\label{dBKP}
\tau\tau_{ijk}-\tau_i\tau_{jk}+\tau_j\tau_{ik}-\tau_k\tau_{ij}=0,\quad
i<j<k.
\end{equation}
The following holds:
\begin{itemize}
\item Eq. (\ref{dBKP}) is four--dimensionally consistent. Moreover, for
the value $\tau_{1234}$ one finds a remarkable equation:
\begin{equation}\label{dBKP1234}
\tau\tau_{1234}-\tau_{12}\tau_{34}+\tau_{13}\tau_{24}-\tau_{23}\tau_{34}=0,
\end{equation}
which essentially reproduces the discrete BKP equation. So,
$\tau_{1234}$ does not actually depend on the values $\tau_{i}$,
$1\le i\le 4$. This can be considered as an analog of the
tetrahedron property of Sect. \ref{sect:theorem}.
\end{itemize}
Notice that usually the discrete BKP equation (\ref{dBKP}) is
written in a slightly different and more symmetric form, with all
plus signs on the left--hand side. On every three--dimensional
subspace these two forms are easily transformed into one another.
However, this cannot be done on the whole of ${\mathbb Z}^4$. Eq.
(\ref{dBKP}) with all plus signs on the left--hand side {\it does
not} possess the property of the four--dimensional consistency.

Further, we mention systems of the geometrical origin (discrete
analogs of conjugate and orthogonal coordinate systems)
\cite{DMS}, \cite{BP2}, which also have the property of the
four--dimensional consistency. We plan to address various aspects
of the four--dimensional consistency in our future publications.
\vspace{5pt}

\paragraph{Acknowledgment.} This research was partly supported by
DFG (Deutsche Forschungsgemeinschaft) in the frame of SFB 288
''Differential Geometry and Quantum Physics''.


\end{document}